\documentclass[twocolumn,letterpaper,amsmath,amssymb,floatfix,aps,superscriptaddress]{revtex4}
\pdfoutput=1

\usepackage{graphicx}
\usepackage{dcolumn}
\usepackage{bm}
\usepackage{epsfig,color}

\usepackage[version=3]{mhchem} 
\usepackage[T1]{fontenc}       

\usepackage[fontsize=10pt]{scrextend}

\usepackage{bm} 
\usepackage{float}
\usepackage{cancel}

\usepackage[hyperfootnotes=false]{hyperref}	 
\hypersetup{
  colorlinks,
  citecolor=blue,
  linkcolor=blue,
  urlcolor=blue
}
\usepackage{titlesec} 
\titleformat{\subsection}[runin]
{\normalfont\bfseries}{\thesubsection{.}}{1em}{}[.]

\titleformat{\section}
{\normalfont\sffamily\large\bfseries}{\thesection{.}}{1em}{\MakeUppercase}

\usepackage{graphicx}
\usepackage{amsmath,fullpage,amssymb}
\usepackage{graphics,epsfig}
\usepackage{makeidx}
\usepackage{gensymb}
\usepackage{multirow, array}
\usepackage{ulem}

\def\ln{{\operatorname{ln}}}

\def\rmd{{\mathrm{d}}}

\def\rme{{\mathrm{e}}}

\newcommand{\Av}[1]{{\bf #1}}

\newcommand{\kB}{k_\mathrm{B}}

\usepackage{soul}  

\begin{document}

\title{Interaction of charged patchy protein models with like-charged polyelectrolyte brushes}

\author{Cemil Yigit}
\affiliation{Institut f\"ur Physik, Humboldt-Universit\"at zu Berlin, 12489 Berlin, Germany}
\affiliation{Institut f\"ur Weiche Materie und Funktionale Materialien, Helmholtz-Zentrum Berlin, 14109 Berlin, Germany}
\affiliation{Helmholtz Virtual Institute, Multifunctional Biomaterials for Medicine, 14513 Teltow, Germany}
\author{Matej Kandu\v{c}}
\affiliation{Institut f\"ur Weiche Materie und Funktionale Materialien, Helmholtz-Zentrum Berlin, 14109 Berlin, Germany}
\author{Matthias Ballauff}
\affiliation{Institut f\"ur Physik, Humboldt-Universit\"at zu Berlin, 12489 Berlin, Germany}
\affiliation{Institut f\"ur Weiche Materie und Funktionale Materialien, Helmholtz-Zentrum Berlin, 14109 Berlin, Germany}
\affiliation{Helmholtz Virtual Institute, Multifunctional Biomaterials for Medicine, 14513 Teltow, Germany}

\author{Joachim Dzubiella}
\affiliation{Institut f\"ur Physik, Humboldt-Universit\"at zu Berlin, 12489 Berlin, Germany}
\affiliation{Institut f\"ur Weiche Materie und Funktionale Materialien, Helmholtz-Zentrum Berlin, 14109 Berlin, Germany}
\affiliation{Helmholtz Virtual Institute, Multifunctional Biomaterials for Medicine, 14513 Teltow, Germany}
\email{joachim.dzubiella@helmholtz-berlin.de}

\pagenumbering{arabic}
\noindent

\parindent=0cm
\setlength\arraycolsep{2pt}

\begin{abstract}
We study the adsorption of charged patchy particle models (CPPMs) on a thin film of a like-charged and dense polyelectrolyte (PE) brush (of 50 monomers per chain)  by means of implicit-solvent, explicit-salt Langevin dynamics computer simulations. Our previously introduced set of CPPMs embraces well-defined one-, and two-patched spherical globules, each of the same net charge and (nanometer) size, with mono- and multipole moments comparable to those of small globular proteins. We focus on electrostatic effects on the adsorption far away from the isoelectric point of typical proteins, i.e., where charge regulation plays no role. 
Despite the same net charge of the brush and globule we observe large binding affinities up to tens of the thermal energy, $\kB T$, which are enhanced by decreasing salt concentration and increasing charge of the patch(es). Our analysis  of the distance-resolved potentials of mean force together with a phenomenological description of all leading interaction contributions shows that the attraction is strongest at the brush surface, driven by multipolar, Born (self-energy), and counterion-release contributions, dominating locally over the monopolar and steric repulsions.  
\end{abstract}
\maketitle

\thispagestyle{plain}
\pagestyle{plain}

\pagenumbering{arabic}
\setlength\arraycolsep{2pt}

\section{Introduction}

Polyelectrolyte brushes consist of polyelectrolyte chains grafted to a planar or curved surface~\cite{Ruhe2004,Ballauff2007:1,Minko_2006}.
The overall structure of such a system is mainly determined by the osmotic pressure of the counterions. In the osmotic
limit, that is, at low salt concentrations, the chains are strongly stretched whereas in the salted brush generated by
high salt concentrations the electrostatic interaction is strongly screened and the resulting spatial structure of the brush
layer resembles the case of uncharged brushes~\cite{Pincus1991:1, Zhulina1995:1, Csajka2000:1, sim2, sim3, sim4, sim5, Ballauff2006:1}. In this way
polyelectrolyte brushes present adaptive systems that have been discussed for a broad variety of applications~\cite{Minko_2006}.
Some 10 years ago it has been found that polyelectrolyte brushes strongly adsorb proteins with like net charge in the osmotic
limits, while virtually no interaction took place in the limit of the salted brush~\cite{Wittemann2003:1}. Thus, a spherical
polyelectrolyte brush consisting of a solid core of approximately 100~nm diameter carrying long grafted chains of poly(acrylic acid) (PAA) adsorbs high amounts of bovine serum albumine (BSA) above its
isoelectric point where the overall charge of the protein is negative as well. This discovery of the adsorption of proteins
at the ``wrong side'' of the isoelectric point has led to a number of experimental and theoretical studies on planar and
spherical systems~\cite{Biesheuvel2005:1, Wittemann2006:1, Biesheuvel2006:1, Dai, Kusumo, Leermakers2007:1, Becker2011:1, Yang}.

In principle, the brush and the protein should repel each other because of two reasons, namely i) electrostatic
repulsion, and ii)~the electrosteric repulsion between the protein and the brush layer: Inserting a protein into a polyelectrolyte brush will lead to unfavorable steric interactions with tethered chains of the brush as well as to a raise of the osmotic pressure of the confined counterions. Hence, there must be a strong attractive force which is capable of overcoming these strongly repulsive  forces.
Three major causes have been suggested to explain the strong attraction observed for a broad variety of systems:
 
 \begin{itemize}
	 \item  First, charge inversion of the protein immersed in a polyelectrolyte brush~\cite{Biesheuvel2005:1,Biesheuvel2006:1}. The pH-dependent protein concentration within the brush layer my be lower than outside in the bulk and below the isoelectric point. Hence, the net charge of the protein changes its sign.
	  
\item As a second driving force counterion release was invoked~\cite{Record1978:1,Czeslik2004:1, Wittemann2006:1}.  Proteins in general carry patches of positive charge on their surface even above the isoelectric point. These patches can interact with the negatively charged polyelectrolyte
chains, thereby releasing a certain number of previously bound counterions into the bulk phase. Since the osmotic pressure within the brush
layer at low salt concentration is quite high, it has been argued that the effect of counterion release should be strong and
virtually independent of the proton concentration within the brush~\cite{Wittemann2003:1,Henzler2010:1}.
\item As a third cause for attraction, a 
heterogeneously charged protein globule may be attracted into the brush because the polyelectrolyte brush interacts asymmetrically 
with the dipole of the protein caused by large charge patches of opposite sign~\cite{Leermakers2007:1,deVos2008:1,deVos2009:1,deVos2010:1}.

\end{itemize}

\textcolor{black}The combined problem 
of patchiness and counterion release has been re-considered by de Vos {\it et al.} in a field-theoretical study~\cite{deVos2010:1}.  These authors concluded that the effect of counterion release is
operative but of minor importance. However, a recent study by He et al.~\cite{He_Merlitz_Sommer_Wu_2015} using molecular dynamics simulations clearly underscored the important role of counterion release when considering the interaction of proteins with polyelectrolyte brushes. Thus, the role of counterion release and its magnitude is still not fully elucidated.

In order to re-consider this problem, we first developed a model for proteins with patchy
surface charge (charged patchy protein model, CPPM)~\cite{Yigit2015:2}. The interaction of these model proteins with single polyelectrolyte
chains of like charge was studied extensively by implicit-solvent/explicit-salt Langevin computer simulations~\cite{Yigit2015:1, Yigit2015:2}. The results obtained in this study confirmed the general concept of counterion release for highly charged polyelectrolytes. Moreover, a simple theoretical model was shown to capture the salient features of the simulations. In particular, the binding affinity derived from these simulations using the CPPM-model was shown to scale dominantly with the logarithm
of the salt concentration in the bulk multiplied by the number of released counterions~\cite{Yigit2015:1}. This relation had been predicted by Record and Lohman~\cite{Record1976:1} a long time ago and found to be valid in a number of experimental studies~\cite{Rau1992:1, Yu2015:1}.
It demonstrates the direct relation between the binding affinity and the translational entropy of the released counterions. It can be derived directly for the problem under consideration here as shown by Henzler {\it et al.}\cite{Henzler2010:1}.

Here we extend our previous studies~\cite{Yigit2015:1, Yigit2015:2} to explore the interaction of patchy proteins
with a thin planar film of a dense like-charged polyelectrolyte brush. In order to keep the problem as simple as possible, we do not consider the effect of charge
reversal and focus on the electrostatic mechanisms for fixed (pH-independent) local charge distributions. 
The general goal of this study is a fully quantitative assessment of the local electrostatic effects that drive like-charged attraction with a sharpened  focus on the details of dipolar and counterion-release effects in the limit of highly charged PE brushes. Our simulations allow us to accurately average out all conformational effects and 
explore in detail the ionic mechanisms taking place at the PE-patch interface. Based on this, the various contribution to protein uptake by the brush  can be properly characterized and quantitatively compared to analytical models. 

\begin{table*}[htb]
 \caption{A summary of our charged patchy particle models (CPPMs) denoted by $P_s^m$. The index $m$ stands
 for the  number of patches, while $s$ denotes the number of positive charges on each patch. In the images of
 the CPPMs in  the top row, the pink beads depict the negatively charged atoms, while turquoise beads depict
 the positively ones. Yellow and white atoms depict the same neutral atoms and are only distinguished here to
 better illustrate the patch region, which roughly has an area of 3~nm$^2$. All CPPMs have a radius of $R_\textrm P=2$~nm
 and a net charge of $Q_\textrm P=-8$~e. The patchy globules carry individual dipole moments as also summarized in the
 table. The corresponding quadrupole (tensorial) moments are provided in previous work~\cite{Yigit2015:1}.}
 \begin{tabular}{l|c|c|c|c|c|c}
 \vspace{-1mm}               & \includegraphics[width=0.09\textwidth]{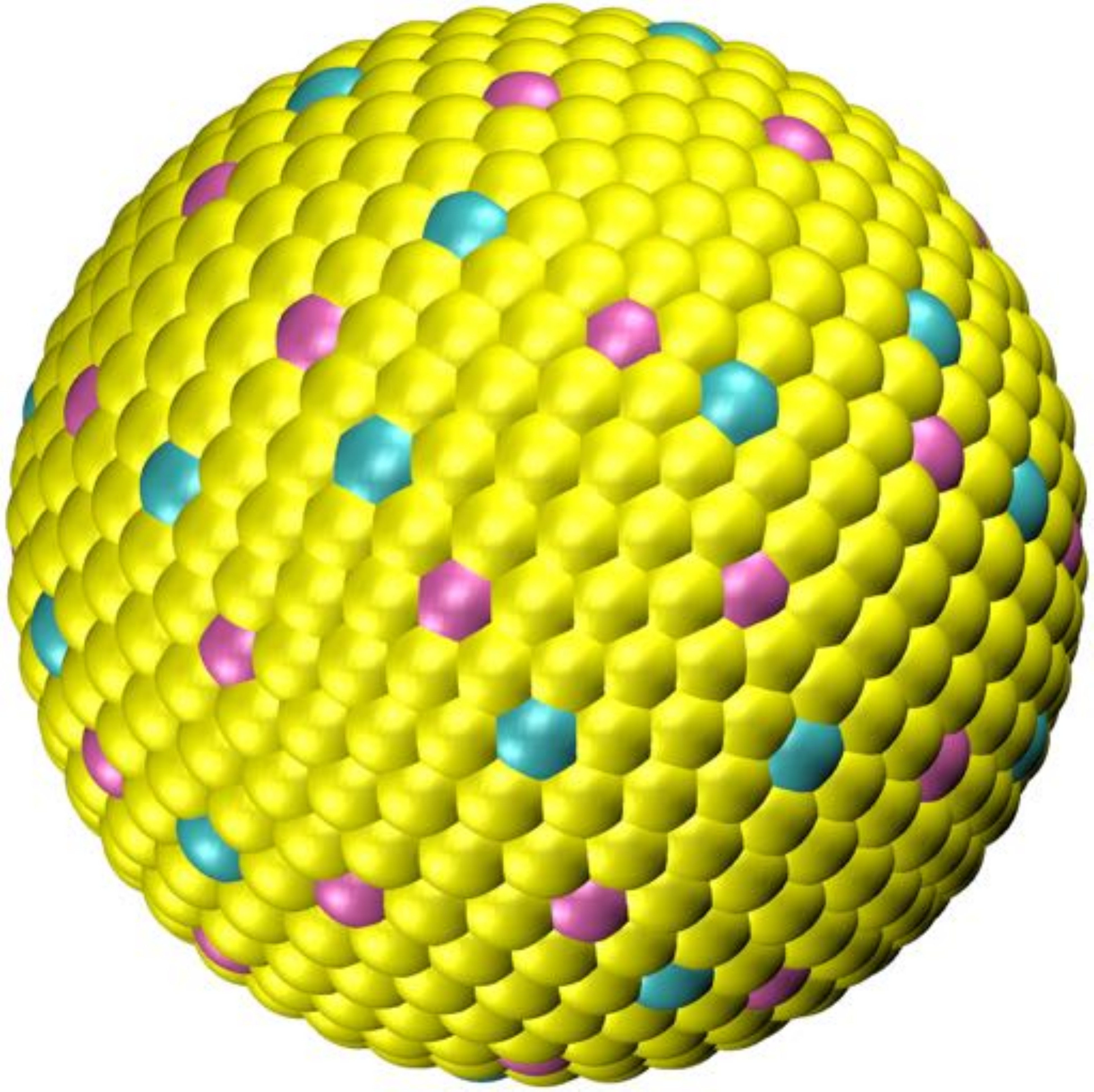}
                              & \includegraphics[width=0.09\textwidth]{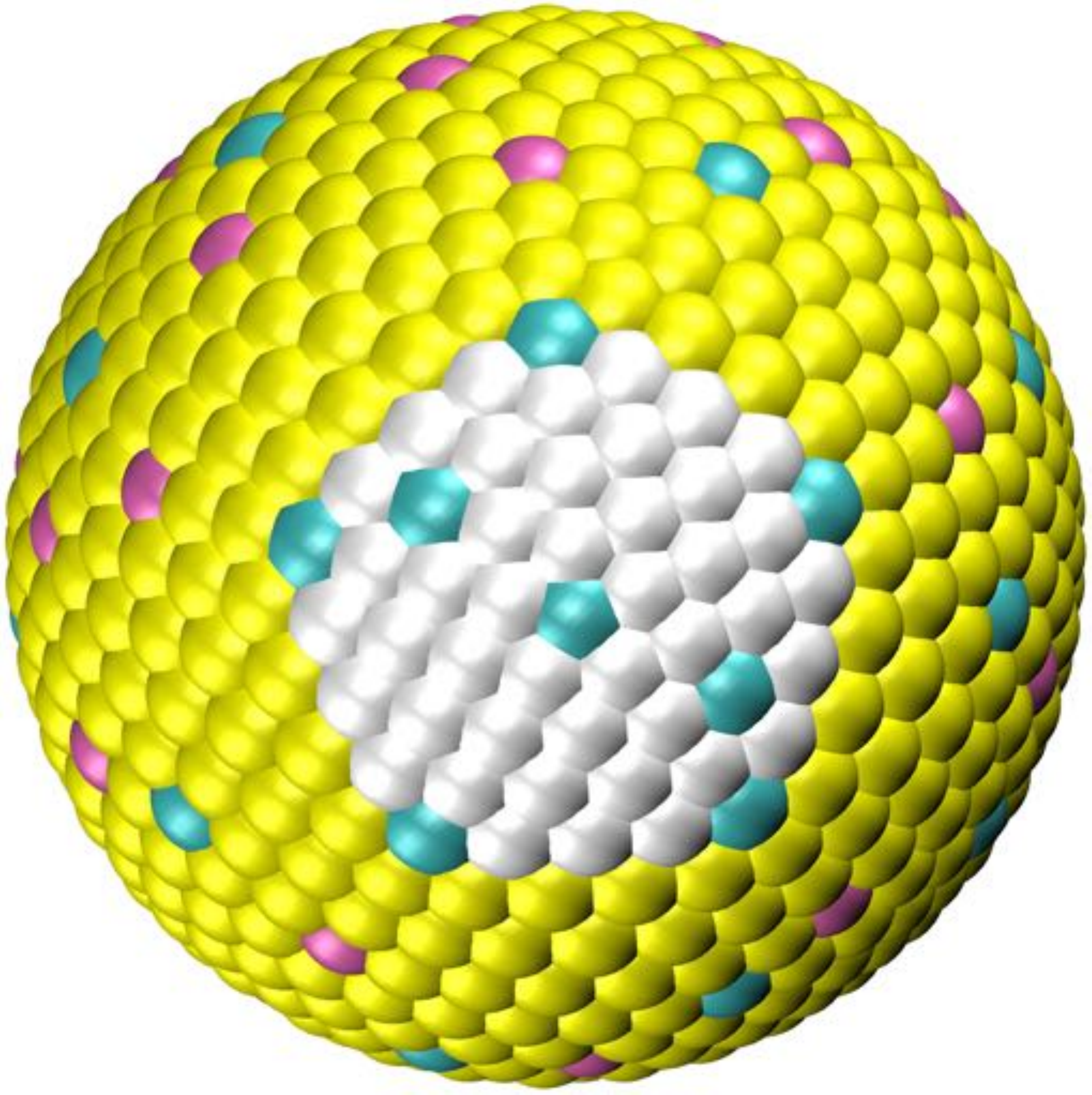}
                              & \includegraphics[width=0.09\textwidth]{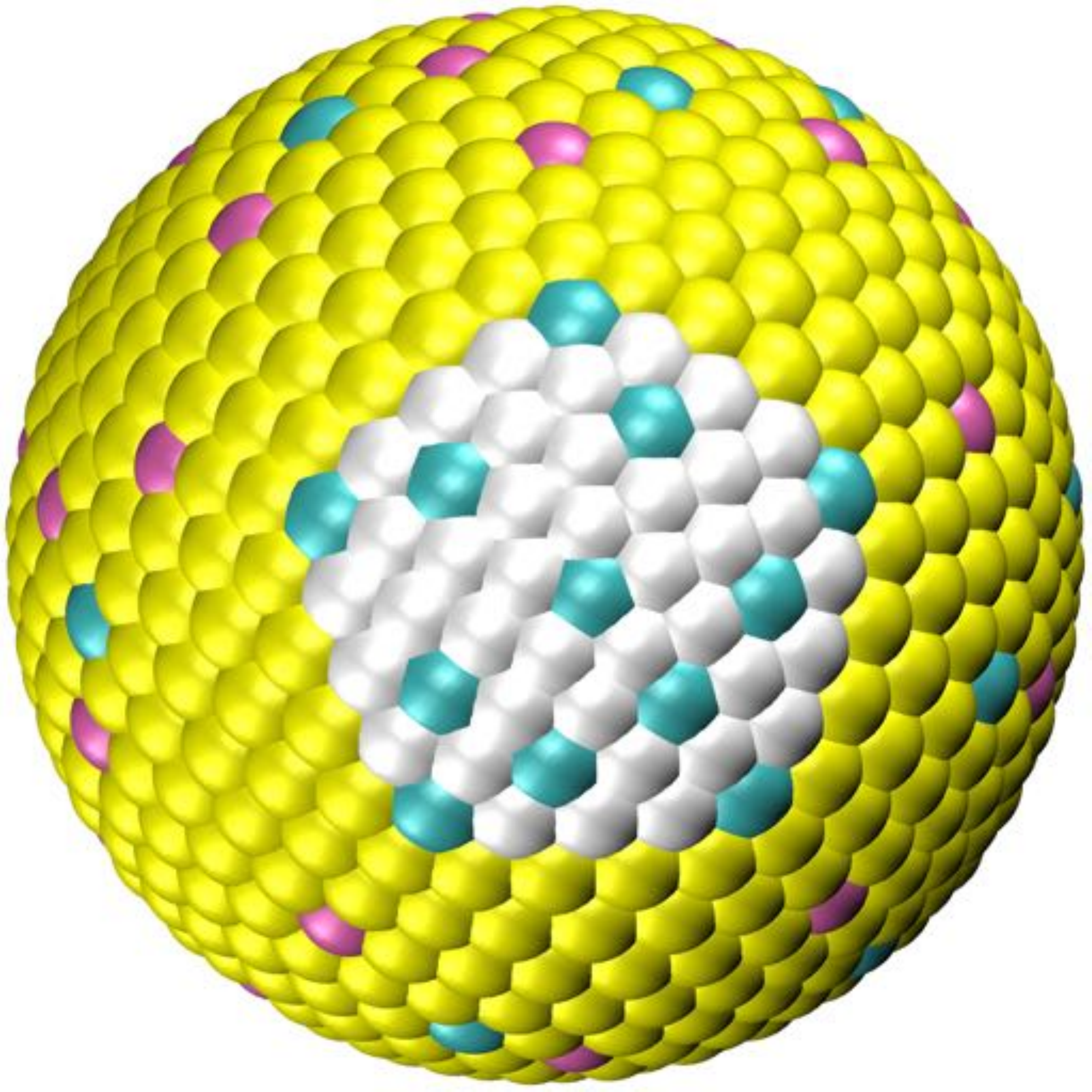}
                            & \includegraphics[width=0.09\textwidth]{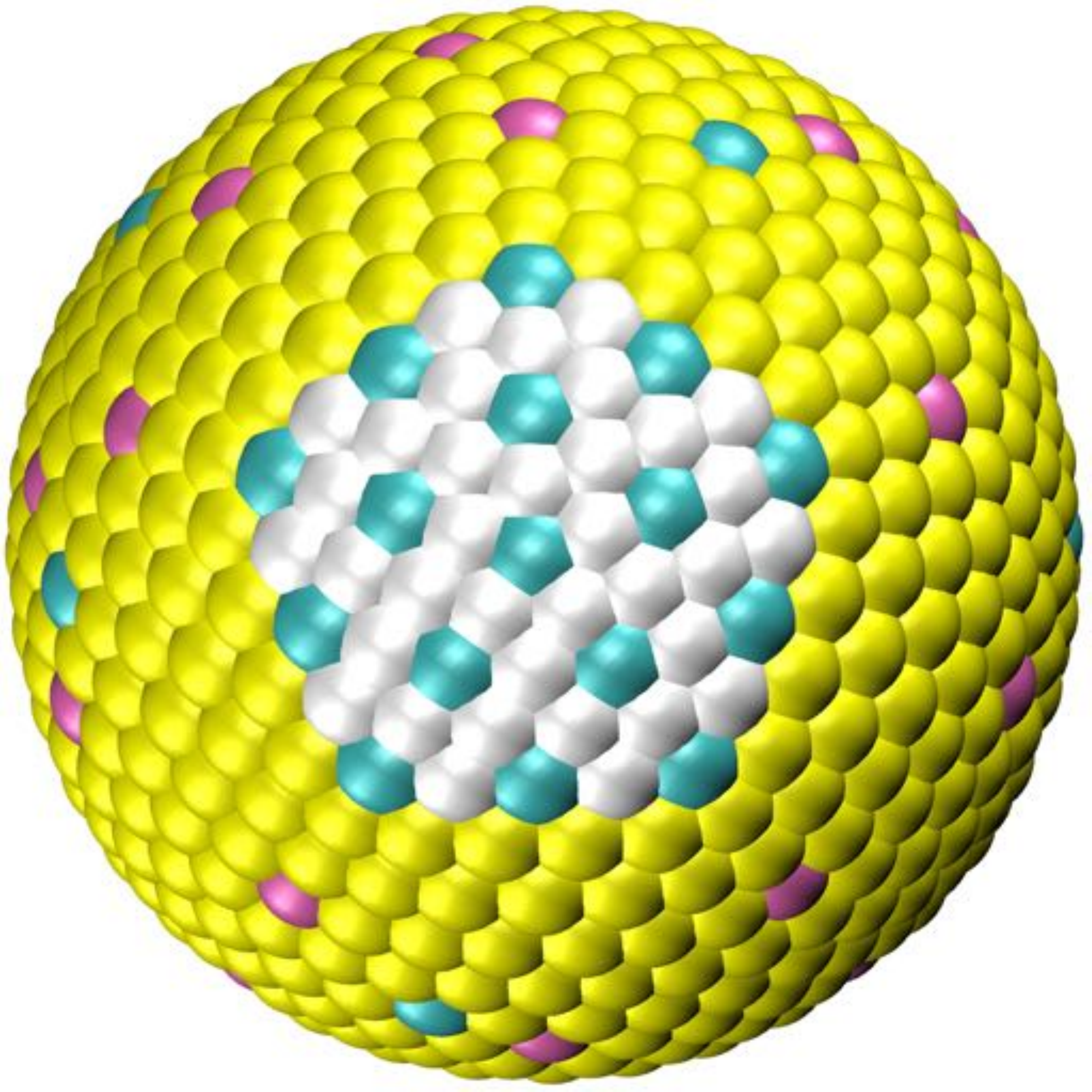}
                            & \includegraphics[width=0.09\textwidth]{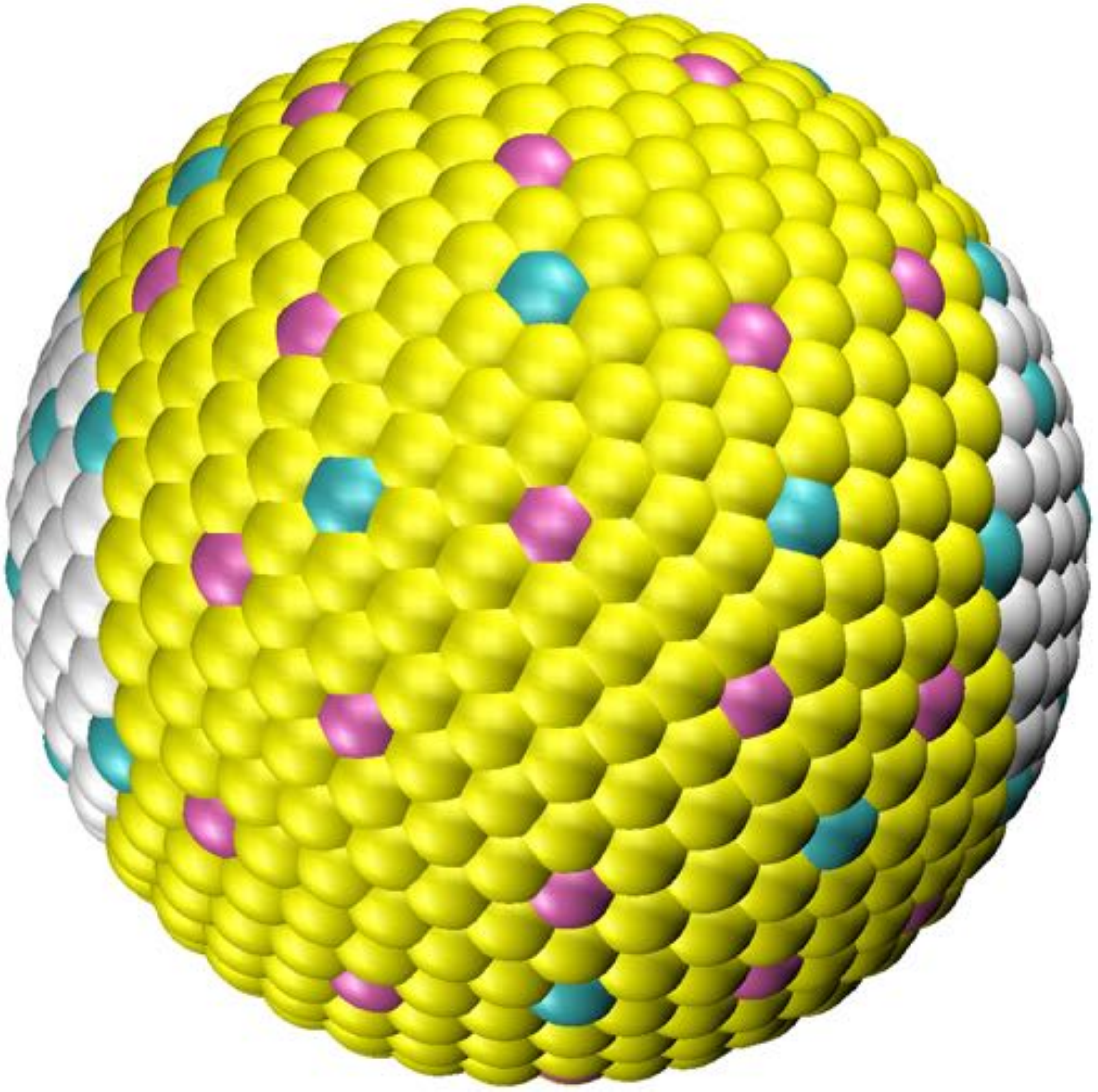}
                          & \includegraphics[width=0.09\textwidth]{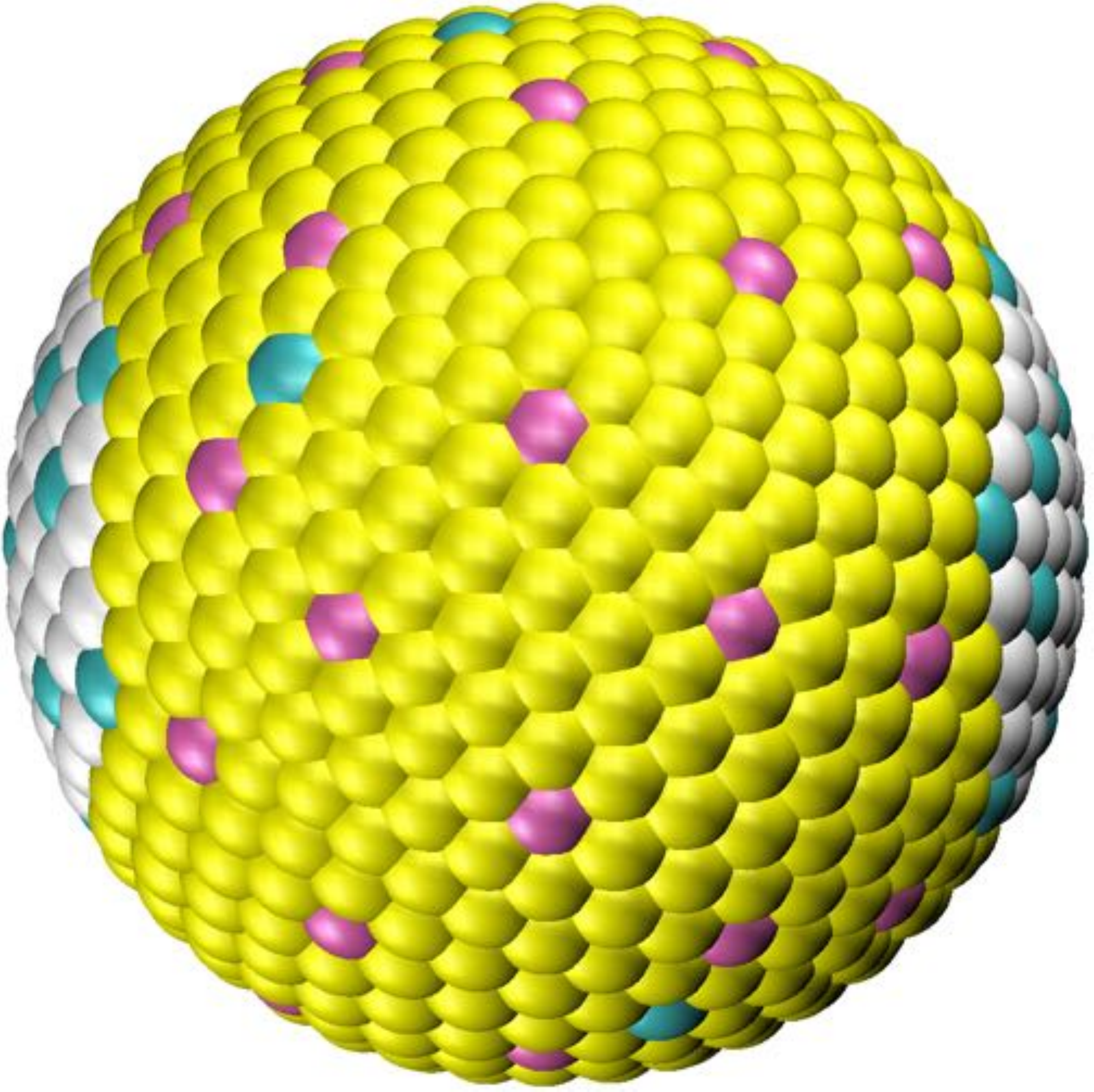}\\\hline\hline
  Label                       & $P^0_0$ & $P^1_8$ & $P^1_{12}$ & $P^1_{16}$ & $P^2_{8}$        & $P^2_{12}$\\\hline
  Radius $R_\textrm P$ [nm]           & 2       & 2       & 2          & 2          & 2                & 2\\\hline
  Patch area $A_\textrm P$ [nm$^2$]   & 0       & 3       & 3          & 3          & 3 (x2)           & 3 (x2)\\\hline
  Total charge $Q_\textrm P$ [e]      & -8      & -8      & -8         & -8         & -8               & -8\\\hline
 Dipole moment $\mu_\textrm P$ [D]   & \textcolor{black}{159}  & \textcolor{black}{896}  & \textcolor{black}{1329}    & \textcolor{black}{1633}    & \textcolor{black}{206}           & \textcolor{black}{151}\\\hline\hline
 \end{tabular} 
 \label{tab:cppm}
\end{table*}

We systematically start by studying the interaction of uncharged spheres with the brush layer to obtain the simplest steric interaction. In the second step, the interaction of a CPPM without patches but a net charge with the like-charged brush layer is determined to study exclusively the repulsive
electrostatic interactions based on monopole interactions and the contributions by the osmotic pressure of the counterions.  Finally, we study CPPMs with additional charge patches (with opposite sign of the charge than the brush) that indeed can display a strong, like-charge attraction. We qualitatively discuss the balance between all leading-order interaction mechanisms
at hand of a phenomenological brush--CPPM binding model.   As in our previous paper,\cite{Yigit2015:1} we find that the 
counterion-release effect (involving 2--3 ions) is one of the dominant driving force for protein adsorption. However, this interaction can be  supported significantly
by the dipolar attraction and Born (self-energy) terms. We demonstrate that the combination of all terms lead to a stable adsorption of the protein at the surface of the brush layer.

\section{Methods}
\subsection{Charged patchy protein models (CPPMs)}

As introduced earlier,~\cite{Yigit2015:2, Yigit2015:1}  we employ a set of spherical patchy protein models 
with well-defined charge patchiness and multipolarity to mimic the electrostatic features of nanometer-sized globular proteins 
or similar nanoparticles.  Briefly, the CPPMs are constructed by distributing 642 atom-sized ($\simeq 0.3$ nm) 
beads on a spherical  surface with radius $R_\textrm P=2$~nm. The latter is typical for small globular proteins such as lysozyme or lactoglobulin~\cite{Uversky1993:1}. 

To build a charge patch, one bead on the surface is randomly chosen and subsequently the closest neighboring
beads are selected until a roughly circular patch area $A_\textrm P \simeq 3$~nm$^2$ is achieved. This area is of
the same order as the size of some naturally occurring larger clusters of charged amino acids of the same sign,
based on the inspection of crystal structures of small globular proteins~\cite{Berman2000:1}.
Afterwards, $s$ positive charges are placed on randomly chosen beads on the patch. We construct  patchy globules with one $(m=1)$ or
two $(m=2)$ patches. In CPPMs with two patches the patches are antipodally directed, that is, on
the exact opposite sides. In order to assign a net charge $Q_\textrm P$ to the CPPMs, we fixed the number
of negatively and positively charged beads to be $N_n=37$ and $N_p=29$ in all CPPMs. Thus, the
resulting net charge of the patchy globule is $Q_\textrm P=-8$~e for all CPPMs, comparable to absolute net
charges of proteins of similar size at physiological conditions~\cite{Berman2000:1}. The $N_n$ negative charges are homogeneously
distributed on the surface around the positive patch(es). The remaining $N_p-m\cdot s$ positive charges are
distributed in such a way that charged beads (positive or negative) are not immediately adjacent. Our
CPPMs are denoted by $P^m_s$ where $m$ specifies the number of patches and $s$ the number
of positively charged beads per patch.

Illustrative snapshots and a summary of the CPPM features, in particular, the dipole moments, are listed in
Table~\ref{tab:cppm}. In our models we consider $m=1,2$ and $s=8,12,16$ resulting in mean patch charge densities
of around 1 to 2~e/nm$^2$ corresponding to a local assembly of a few amino acids separated from
each other by a few angstroms~\cite{Kayitmazer2013:1, Berman2000:1}. The dipole moments, \textcolor{black}{calculated using the center-of-mass of the CPPM
as coordinate origin for the charge distribution}, are in the
range of 159~Debye to 1633~Debye, \textit{cf.} Table~\ref{tab:cppm}, also comparable to proteins of this size.
The small lactoglobulin, for instance, has 730~Debye~\cite{Ferry1941:1}.
\subsection{PE brush model}

\begin{figure}[t!]
 \centering
 \includegraphics[width=0.45\textwidth]{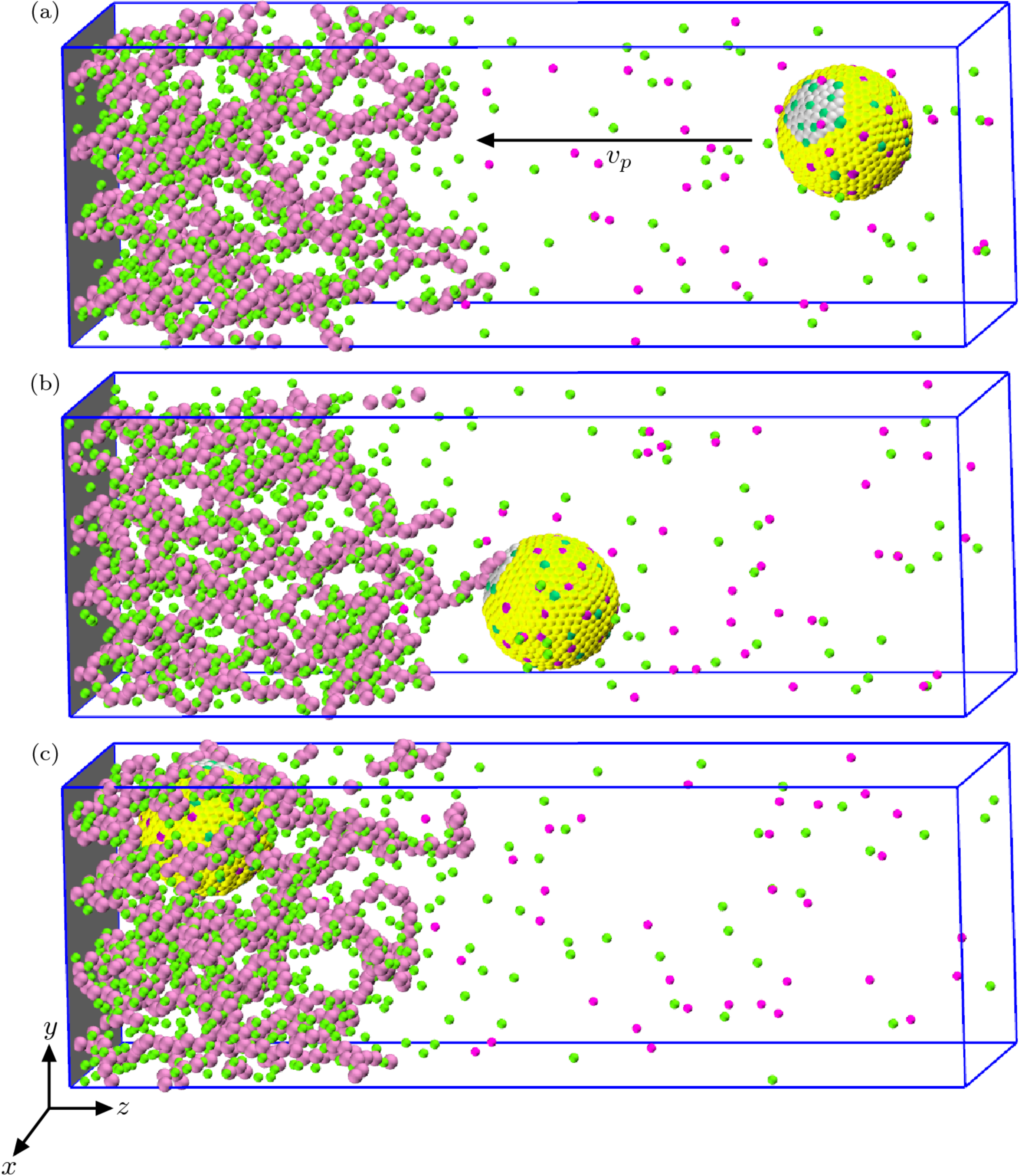}
 \caption{An illustration of the uptake of a CPPM (yellow patchy globule) by a planar PE brush (magenta colored connected beads) 
 in the presence of co- and counterions (small magenta and green beads, respectively). (a) The CPPM is
 situated in the bulk region. (b) for the PMF calculation the CPPM is moved with a constant pulling rate $v_p$ towards the brush and 
 (c) further pulled through the brush layer until it finally reaches the grafting surface.}
 \label{fig:protein_uptake}
\end{figure}

The simulated PE brush  is composed of 16 flexible PE chains fixed by harmonic constraints 
at one end on a neutral and planar surface in equidistant spacings on a square lattice. With a surface area of 100~nm$^2$, the grafting density thus corresponds
to $\tau_B=0.16$~${\text{molecules}}/{\text{nm}^2}$. A single flexible PE is modeled in a coarse-grained
fashion as a sequence of $N_\text{mon}=50$ freely jointed beads. Each bead represents a monomer with a diameter
$\sigma_{\rm LJ}$ and an electric charge of one negative elementary charge $-e$. 
 The PE monomers are connected by a harmonic bond potential with an equilibrium bond length
$b_\text{mon}=0.4$~nm and a force constant $K_\text{mon}=4100$~${\text{kJ}}/{\text{(mol nm}^2)}$. A harmonic 
angle potential is applied in which the angle between a triplet of
monomers is $\gamma=120^\circ$ and the force constant is $K_\gamma=418$~${\text{kJ}}/{\text{(mol rad}^2)}$.
 An illustrating snapshot of the PE brush (including salt) and a CPPM is shown in Fig.~\ref{fig:protein_uptake}.
\subsection{Simulation details}
The dynamics of each of the beads (creating the CPPM and the brush) and explicit ions is governed  by Langevin's equation of motion
\begin{equation}
 m_i\frac{d^2\bm{r}_i}{dt^2} = -m_i\xi_i\frac{d\bm{r}_i}{dt} + \bm\nabla_{i}U + \bm{R}_i(t)
\end{equation}
where the force $\bm{R}_i(t)$ is a Gaussian noise
process with zero mean $\langle \bm{R}_i(t)\rangle =0$ and satisfies the fluctuation--dissipation theorem
\begin{equation}
 \langle \bm{R}_i(t) \cdot \bm{R}_j(t') \rangle = 2m_i\xi_i\kB T\delta(t-t')\delta_{ij}.
\end{equation}
The $m_i$ and $\xi_i$ are the mass and friction constant of the $i$th bead, respectively. $U$ is the
system potential energy and includes intra- and intermolecular interactions and position constraints.   
All interatomic interactions are composed of the Lennard-Jones~(LJ) potential
 between the beads, ions, and the grafting surface (modeled by the 9--3 potential), 
 as well as the Coulomb interaction between all 
charged beads and ions.  LJ interactions between neighboring polymer beads are excluded. 
The solvent is modeled as a continuous medium with a static dielectric constant $\epsilon= 78.44$. 
All beads and ions have mass $m_i = 1$~amu, a LJ diameter $\sigma_{\rm LJ} = 0.3$~nm,
energy well $\epsilon_{\rm LJ} = 0.1$~k$_\text{B}$T,  and integer charges $q_i$ = 0, $+e$, or $-e$. 

The simulations are performed using the GROMACS 4.5.4 software package~\cite{Hess2008:1}. A leap-frog
algorithm with a time step of 2~fs is used to integrate the equations of motion. The Langevin thermostat
with $\xi_i = 0.5$~ps$^{-1}$ keeps the temperature at $T = 298$~K and generates a canonical
ensemble ($NVT$). Center of mass translation of the system is removed every 10 steps. The rectangular
simulation box with $L_x=L_y=10$~nm, and $L_z=30$~nm is periodic only in the $x,y$-directions. 
The walls, placed at $z=\sigma$ and $z=30$~nm, are represented by the 9--3 LJ potential
with parameters $\sigma=0.3$~nm, $\epsilon=0.1$~$\kB T$, and parameter 'wall-density' = 0.5 \textcolor{black}{nm$^{-3}$}. One end of the PE chains is position-restrained in the immediate vicinity of the wall at $z=0$~nm
by a harmonic potential with a force constant of 4100~${\text{kJ}}/{\text{(mol nm}^2)}$.  The cut-off radius is set to 1.2~nm to calculate the real-space interactions, while Particle-Mesh-Ewald (PME) is implemented to account for long-range electrostatics~\cite{Essmann1995:1}. The reciprocal summation of the PME method is computed on a 3D FFT grid but with spacings of 0.32~nm in $x,y$-directions and 0.23~nm in $z$-direction using a fourth-order
interpolation. Because of the periodicity, a correction term to the Ewald summation in the $z$-direction
is added to produce a pseudo-2D summation~\cite{Yeh1999:1, Qiao2011:1} 
\textcolor{black}{to avoid artefacts of the system's instantaneous net dipole moment.}.

The CPPM is initially placed at $z\approx27$~nm. After the simulation box with brush and CPPM is set
up, the corresponding number of counterions is added to ensure electroneutrality of the system. Additionally,
monovalent salt is added to the system leading to (bulk) concentrations $c_s = 15, 32, 78, 137$, and $259$~mM. 
The system is relaxed for 100~ps to remove local contacts and afterwards equilibrated for 30~ns.
\subsection{PMF calculations}
For calculating the PMF between the brush and CPPMs the pull code of the software package GROMACS was used with
the umbrella method~\cite{Hess2008:1}. Here, the center-of-mass (COM) of the patchy particle is restrained in space by an
external time-dependent force. This force is applied by a harmonic potential that is 
moved with a constant pulling velocity $v_p$ to steer the CPPM in the prescribed direction~\cite{Isralewitz2001:1}.
The reaction coordinate is $z$, the distance of the COM of the CPPM to the grafting surface at $z=0$. 
After several test runs, and comparison to standard umbrella sampling,~\cite{Yigit2015:1} the  pulling rate $v_p = 0.1$ nm/ns 
and a harmonic force constant $K = 2500$ kJ mol$^{-1}$ nm$^{-2}$ were chosen to yield the best performance with respect to 
accuracy versus computational effort.  The simulation time of $\sim$250~ns is required to pull the CPPMs deep into the brush ($z\simeq 2$~nm) from a  separated state ($z \simeq 27$ nm). The standard deviation was calculated by block averages to specify the statistical error. 

After a completion of a run, the constant friction force $F = -m\xi v_p$ was subtracted from the total 
force and averaged within a specific interval of discrete spacing $\Delta z$ to obtain the mean force.
According to our simulation setup, the mean force was integrated backwards to get the PMF. 
We emphasize that in all our simulations the patchy particles were able to rotate freely and, thus,
all our results are orientation-averaged with the appropriate and correct Boltzmann weight.

\subsection{CPPM orientation}

 The patch vector $\bm{p}$
points from the particle center to the patch center and provides also a very good approximation
of the dipole direction $\vec \mu$ of the P$^1_s$ models. In our analysis we computed the distance-resolved
cosine of the angle $\theta(z)$ by
\begin{equation}
 \cos [\theta(z)] = \left\langle \frac{\bm{p}\cdot\bm{e}_z}{\lvert\bm{p}\rvert}\right\rangle_z \quad.
 \label{eq:cosines}
\end{equation}
The distance-dependent angular correlation of the patch vectors is calculated via the second Legendre polynomial $P_2(\cos[\theta])$ with $P_2(x) = (3x^2-1)/2$. For CPPMs with two patches only one patch is used to calculate the
orientation since the patches are antipodally directed.

\subsection{Phenomenological separation of the PMF}

In order to describe the dominant physical contributions to the PMF between the single CPPM 
and the brush, we employ a phenomenological approach where we divide up the total free energy
in three major contributions, via
\begin{equation}
 w_\text{tot}(z) = w_\text{excl+vdW}(z) + w_\text{elec}(z) + w_\text{cr}(z).
  \label{eq:W_tot}
\end{equation}
With the first term $w_\text{excl+vdW}(z)$ we represent all contributions for a
{\it neutral} globule, that is, the excluded-volume (including steric and osmotic) 
and van der Waals (vdW) interactions between a totally neutral CPPM (no bead is charged) and the charged PE brush. As a full analytical theory for this part is difficult to develop, we will explicitly 
calculate the PMF between a neutral CPPM interacting with a charged brush in the simulation. 
However, qualitatively the contributions to $w_\text{excl+vdW}(z)$ can be discussed as follows.  

In our simulations, the vdW interactions are modeled by the attractive part of the LJ interaction. This interaction is small because of our choice of the $\epsilon_{\rm LJ}$-parameter of only 0.1 $\kB T$.  The excluded volume 
part originates on one hand from the \textcolor{black}{configurational} response of the PEs to the CPPM excluded volume and on the other hand from 
the osmotic pressure of confined counterions, which also penalizes the intrusion of the globule. 
Based on classical laws of the osmotic pressure of semi-dilute polymer solutions~\cite{deGennes}
the former is expected to be repulsive on the order  of several $\kB T$ \textcolor{black}{due to the polymer configurational entropy loss}. We will calculate it by simulating a neutral CPPM interacting with a completely neutral brush. Note, however, that charging the polymers stretches the brush and  may modify this  contribution, likely decrease it slightly as the brush effectively swells upon charging. 
The second excluded volume contribution,  let us write it as $w_\text{osm}(z)$, arises from the volume 
work of the penetrating CPPM with volume $V_P$ against the osmotic pressure of the confined  
counterions. Assuming ideal gas behavior, we can expect it to scale roughly linearly with 
counterion density as  $w_\text{osm}(z) = \kB T [c_\text{free}(z) - c_\text{s}] \cdot V_\textrm P$,
where $c_\text{free}(z)$ denotes the local number density of free (not condensed) counterions within the PE brush, while $c_\text{s}$ is the bulk number density of salt ions. For our models, $V_\textrm P \simeq 33.5$~nm$^3$, and 
for a brush charge density of about 1 nm$^{-3}$, then  tens of the thermal energy $\kB T$ repulsion are easily possible.  We will indeed see that  this contribution  dominates the non-electrostatic interaction. 

If the CPPM is charged, explicit electrostatic interactions come into play, incorporated in $w_\text{elec}(z)$ in eq.~(\ref{eq:W_tot}). In principle one could employ numerical solutions of mean-field theories~\cite{Leermakers2007:1}.   
Here we proceed on a simpler but analytical and thus more transparent way: based on analytical solutions of the linearized Poisson--Boltzmann equation  in a cell-model,~\cite{Yigit2012:1} the leading order electrostatic interaction energy for a  strictly monopolar globule  within 
a homogeneous PE matrix is given by two terms: first, a monopole term that simply describes 
charge repulsion by the electrostatic (Donnan) potential. Secondly, a Born term \textcolor{black}{that describes the change of the self-energy of electrostatic charging the globule in the bulk solvent versus the brush environment. The latter  has significantly different  screening properties than the bulk and the self-energy of charging can change considerably.} 
Here, we tie up on these ideas and include the next leading term in  a multipolar expansion of the problem, that is, \textcolor{black}{the dipolar contribution to both the electrostatic interaction~\cite{Hill}} and the Born terms. 
We thus arrive at 
\begin{eqnarray}
 \label{eq:W_elec}
 w_\text{elec}(z) &=& Q_\textrm P\cdot\Phi(z) \\ \nonumber 
   &-&\kB T \;\ln \left[\frac{\sinh[\beta \mu E(z)]}{\beta \mu E(z)}\right] + \Delta w_{\rm Born}(z),
\end{eqnarray}
where $\beta = {(\kB T)}^{-1}$, and the first two terms on the right hand side characterize the (point) monopole and (point) dipole contribution to the direct interaction of the CPPM with the electrostatic \textcolor{black}{potential $\Phi(z)$ and field $E(z)=-{\rm d}\Phi/{\rm d}z$, respectively. The latter} are determined consistently in our work from the charge density profiles through integration of Poisson's equation. $Z_\textrm P<0$ and $\mu_\textrm P>0$ are the net charge and dipole moment of the 
CPPM, respectively.  Note that for our like-charged systems, the monopole-term is repulsive, 
while the dipole term is attractive. 

For the leading order of the Born energy up to the dipole level we derive \textcolor{black}{(see Appendix~\ref{app:1})}
\begin{eqnarray}
w_{\rm Born}(\kappa)&=&\frac{Q_\textrm P^2}{8\pi\epsilon\epsilon_0 R_p}\frac{1}{(1+\kappa R_\textrm P)} \\ \nonumber &+& \frac{3\mu_\textrm P^2}{8\pi\varepsilon\varepsilon_0 R_\textrm P^3}\,\frac{(1+\kappa R_\textrm P)[2+2\kappa R_\textrm P+(\kappa R_\textrm P)^2]}{[3+3\kappa R_\textrm P+(\kappa R_\textrm P)^2]^2}.
\label{eq:F_born_dipole}
\end{eqnarray}
The first term is the classical result  of a monopolar sphere with valence $Z_\textrm P$ and radius $R_\textrm P$ in a salty environment with inverse screening length $\kappa$ in the Debye-H\"uckel approximation~\cite{McQuarrie}. The second term is the equivalent expression for a point dipole with moment $\mu_\textrm P$ centered in the sphere.

Since we are dealing with an inhomogeneous system, i.e., locally varying ion densities perpendicular to the grafting wall,  we apply a local field approximation, that is,  we assume we can apply the Born theory developed for a 
homogeneous system also in the case of la ocal ($z$-dependent) salt distribution.
This should be a good approximation as long as the length scales of the inhomogeneities are larger than the globule size. This is not always the  case in our work but the simple theory will still serve as a good, at least qualitative interpretation of the simulation results.  
Hence, the change of the total Born energy by transferring a CPPM from bulk into the PE brush reads 
in the monopole--dipole approximation
\begin{equation}
  \Delta w_\text{Born}(z) =  w_\text{Born}(\kappa_\text{brush}(z)) -  w_\text{Born}(\kappa_\text{bulk}),
 \label{eq:W_Born_Trans}
\end{equation}
where $\kappa_\text{bulk}  = \sqrt{8\pi\lambda_\textrm B c_s}$  represents the inverse Debye screening length in the bulk and $\kappa_\text{brush}(z) = \sqrt{4\pi\lambda_B \sum_i c_i(z)}$  is the {\it local} inverse screening length within the PE brush. Here, 
the summation $i=+, -, m$ runs over the densities of cations, anions, and PE monomers.
With the latter, we have assumed that the PE monomers are mobile and thus fully contribute to the local screening. Since the PE are flexible and far from close packing, this should be a better approximation than assuming a fixed, non-screening background of the PE matrix. In our comparison to the fully simulated PMFs we will take the salt density profiles $c_i(z)$ directly from reference simulations without CPPM.  
The constant  $\lambda_\textrm B = e^2/(4\pi\epsilon\epsilon_0 \kB T)$ is the Bjerrum length and has the value of 0.71~nm for our systems,  i.e., aqueous solvent at normal conditions.

\begin{figure*}[!htb]
 \centering
 \includegraphics[width=0.8\textwidth]{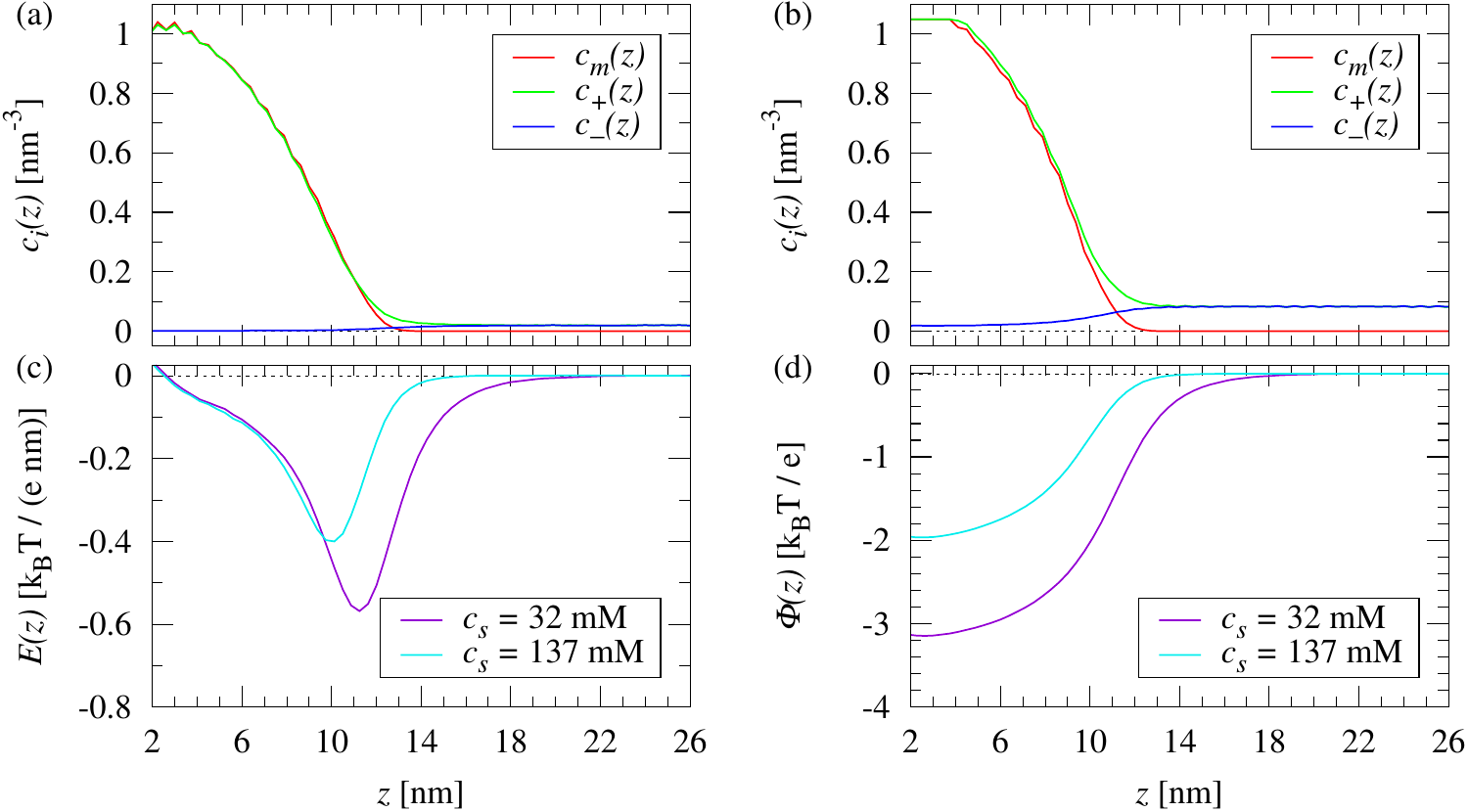}
 \caption{Density profiles $c_i(z)$ for charged PE monomers ($i=m$) and salt ions ($i=\pm$) for an isolated polyelectrolyte brush at (a)~$c_s = 32$~mM and (b) $c_s = 137$~mM. (c) Electrostatic field and (d) potential for a PE brush at 32~mM and 137~mM salt concentration. All PE monomer beads carry a charge of $-e$.}
 \label{fig:isolated_pe_brush}
\end{figure*}

Finally, for our highly charged PE brush interacting with oppositely charged patches, we expect explicit counterion-release effects to play a role, 
which we treat separately to the electrostatic interactions just introduced above. 
In our previous work on the interaction between CPPMs and single PEs,~\cite{Yigit2015:1} we have argued
that counterion-release, within the frame of the Onsager-Manning-Oosawa treatment,\cite{Fuoss1951:1, Oosawa1971:1,Manning1969:1} only happens for ions condensed at the PEs, not those accumulated at the protein patch. After complexation, the positive CPPM patch charges become neutralized by one or more PE chains in the brush and consequently a corresponding amount of  condensed ions will be liberated. 
The release entropy per ion can be of considerable magnitude, that is, of the order of several $\kB T$.  Inspired by the classical work of 
Record and Lohman~\cite{Record1976:1} and more recent work,~\cite{Henzler2010:1} we have shown  that 
the total gain in translational entropy of released ions during CPPM-PE complexation  can be well expressed by~\cite{Yigit2015:1, Yigit2015:2}
\begin{equation}
 w_{cr}(z) = - \kB T N_+(z)\ln\left[\frac{c_{\rm cond}}{c_s}\right], 
 \label{eq:W_CR}
\end{equation}
where $N_+$ is the number of released ions from the PEs upon complexation. As before, $c_s$ is the bulk salt concentration 
and $c_\text{cond}$ (typically $c_{\rm cond} \gg c_s$) specifies the concentration of condensed counterions in the solvation shell  around
the PE monomers. From averaging the radial distribution functions (not shown) in a shell of 0.4~nm thickness we calculate a value 
$c_\text{cond} \simeq 3.5\pm0.5$~M,  only slightly depending on salt concentration~\cite{Yigit2015:1}. 
We refer to previous work for a detailed analysis of accumulation and release of ions during
the association of two CPPMs~\cite{Yigit2015:2}, or a single PE and a CPPM~\cite{Yigit2015:1} or a better resolved protein model~\cite{Yu2015:1}.
In our current work the CPPM complexes with many PE chains at the same time in the brush and the number of released
ions in the simulation is not easily accessible. We circumvent this problem by counting the local number of
PE monomers, $N_{\rm m}(z)$, bound to the patch region and then assuming that on those their fraction of condensed ions were
liberated.  The latter is then estimated using the classical Manning law as $1-\Gamma^{-1}$, 
where $\Gamma = z \lambda_B/l$ is the 'Manning'-parameter for a salt valency $z$ and 
bond length $l\simeq b_{\rm mon}$.  Hence, we find  
\begin{equation}
 w_{cr}(z) = -\kB T \left\{1-\Gamma^{-1}\right\}N_{\rm m}(z)\;\ln\left[\frac{c_{\rm cond}}{c_s}\right]
 \label{eq:W_CR}
\end{equation}
having used the identity $N_+ = \left\{1-\Gamma^{-1}\right\}N_{\rm m}$. For the PEs in our brush we find $\Gamma \simeq 1.78$.

\section{Results}
\subsection{Polyelectrolyte brush only}

We first present an analysis of the monomer and ion density profiles and electrostatic properties of an isolated PE
brush (without any CPPM) at two salt  concentrations.  PE monomer density profiles $c_m(z)$ at ionic strengths of 32~mM and 137~mM are displayed 
in Fig.~\ref{fig:isolated_pe_brush}~(a) and (b), respectively. They are a monotonically decreasing function of the
distance from the grafting surface, quickly converging to zero beyond $z\gtrsim 12$~nm in the bulk region. For the lower 
salt concentration the brush is slightly more stretched, as expected due to the higher osmotic pressure difference of the neutralizing
counterions.  Similar profiles and trends have been observed in related simulations~\cite{Csajka2000:1, sim2,sim3,sim4,sim5} and are predicted by self-consistent field theory~\cite{Zhulina1997:1}. As expected, the counterion profiles $c_+(z)$ closely follow the brush density profiles in the brush layer (apart from a Debye layer at the brush interface at $z\sim 12$~nm), indicating a significant electrostatic neutralization of the PE brush, while the coions are mostly depleted from the brush.
Both the counterion and coion profiles converge to their respective concentration values in the bulk region.

\begin{figure*}[!htb]
 \centering
 \includegraphics[width=0.75\textwidth]{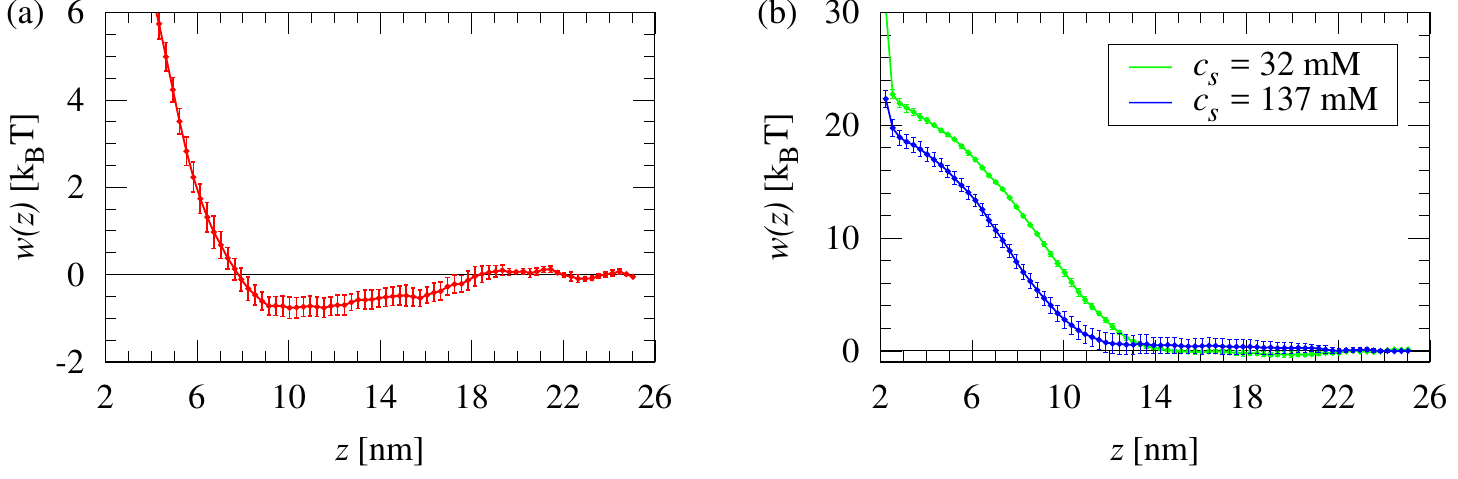}
 \caption{(a) PMF between a fully neutral globule and a neutral PE brush (solid line including error bars) and no salt ($c_s=0$). 
  Panel (b) shows the PMF for a completely neutral globule and charged PE brush at 32 mM and 137~mM salt concentrations.}
 \label{fig:cppm_brush_reference_cases}
\end{figure*}

\begin{figure*}[!htb]
 \centering
 \includegraphics[width=0.75\textwidth]{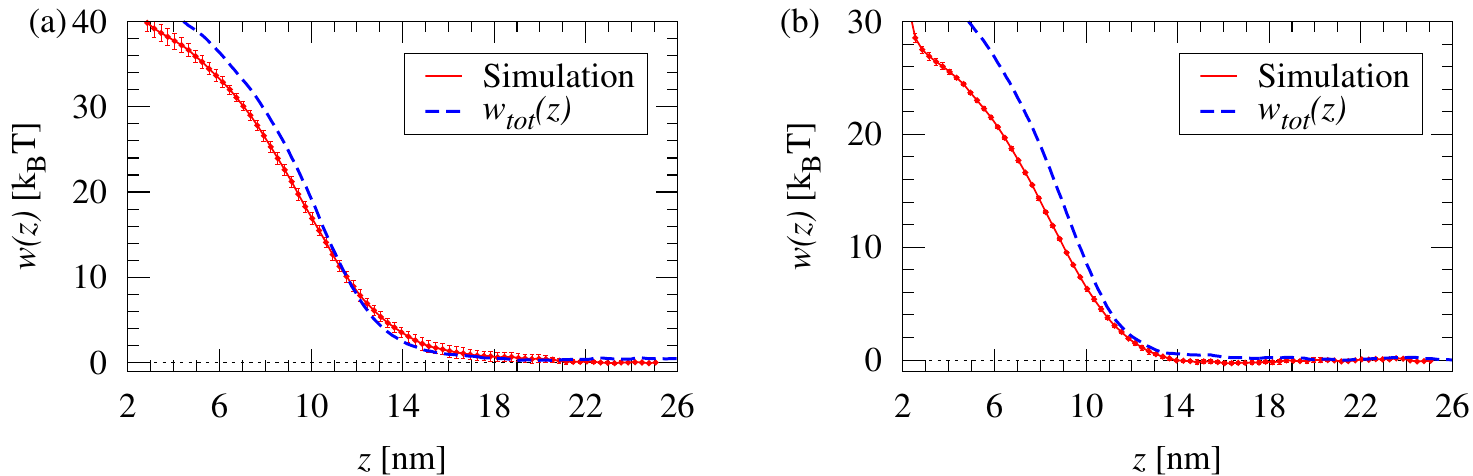}
 \caption{PMF for the charged but patchless CPPM ($P^0_0$) inserted into a charged brush at a salt concentration of (a) 32~mM and (b) 137~mM. The blue dashed lines are comparisons from eqs.(4)--(7) using the potential and ion distributions from the reference simulations in Figs.~2 and 3 as input.}
 \label{fig:cppm_brush_reference_cases_P00}
\end{figure*}

From the charge density profiles we calculate the
electric field $E(z)$ and potential $\Phi(z)$ of the system via integration of Poisson's equation as shown in Fig.~\ref{fig:isolated_pe_brush}~(c) and (d), respectively. The electrostatic potential is negative through the brush layer and saturates to zero in the bulk region. Near the grafting
surface, it reaches roughly $-3.1$~${\text{k}_\text{B}\text{T}}/{\text{e}}$ at the low ionic strength and
about $-2$~${\text{k}_\text{B}\text{T}}/{\text{e}}$ at the higher ionic strength. As a rough consistency check
we can compare these values with the hypothetical Donnan potential that builds up for perfect electroneutralization 
of a charged background connected to a salt reservoir~\cite{Donnan1,Donnan2}. 
The ideal Donnan potential in this case reads $e\beta\phi = \ln (y+\sqrt{y^2+1})$ with $y = c_B/(2 c_s)$, where $c_B$ is the
charge density of the charge matrix (here, the brush). If we take the maximum value of the brush density close to the grafting plane, 
$c_B\simeq 1$~nm$^{-3}$, we find for the electrostatic potentials -3.4 and -2.1 k$_{\rm B}$T/e at the grafting surface for the lower and
higher salt concentration, respectively. These numbers overestimate the simulations by less than 10\%. 
The small discrepancy may come from steric and electrostatic correlation effects, not included in the simple Donnan picture
where ions are simply Boltzmann distributed.

The electrostatic field, $E(z) = -\partial \Phi/\partial z$, plotted in Fig.~\ref{fig:isolated_pe_brush}~(c) shows a distinct
minimum at the brush surface around $z=10-11$~nm, moving to smaller values for the larger salt concentration, when the 
brush shrinks. The reason for this extremum is that here, in the Debye layer, where electroneutrality is locally violated,  
the potential change is extremal before it deteriorates in the inner part of the brush. 
This has interesting consequences on the adsorption of multipolar particles: if a particle has a
strong dipole moment, its dipolar coupling to the field may dominate over other interaction contributions and the
distribution of adsorbed particles may be significantly different than for simple monopoles. 

\subsection{PMFs in some reference cases}

We discuss now a few insightful limiting reference cases of the PMF in order  to understand the
individual interaction contributions better.  In Fig.~\ref{fig:cppm_brush_reference_cases} (a) 
we plot the simplest case where both the brush and the CPPM are completely neutral, i.e., 
no bead is charged and only simple excluded volume and vdW interactions play a role.  
 In the PMF of this neutral system we observe a small 
vdW attraction of about $-1$~k$_\text{B}$T at $z\approx 10$~nm, while for smaller distances the PMF 
is repulsive due to the compression of the brush in a range of a few $k_{\rm B}T$ as 
expected for semi-dilute polymer solutions~\cite{deGennes}.
In Fig.~\ref{fig:cppm_brush_reference_cases} (b) we further plot the PMF of a completely neutral 
CPPM inserted now into a {\it fully charged} brush. In this osmotic brush case the brush is stretched and 
filled with a high density of counterions, cf. Figs.~\ref{fig:isolated_pe_brush}~(a) 
and (b). The repulsion is  considerable in the range of tens of $k_{\rm B}T$, as predicted
from the ideal gas equation. Hence, the contribution $w_\text{excl+vdW}(z)$ 
to the total free energy, eq~(\ref{eq:W_tot}), for a neutral globule pushed into a highly charged brush 
is dominated by the volume work of the globule against the  osmotic pressure of neutralizing counterions. 

As another instructive reference system we plot the PMF of the rather homogeneously charged (patchless) and essentially 
monopolar  ($P^0_0$) system at $c_s=32$~mM and $c_s=137$~mM salt
concentrations in Fig.~\ref{fig:cppm_brush_reference_cases_P00}~(a) and (b), respectively.  
The plots reveal, as intuitively expected, a purely repulsive interaction, weaker for the higher ionic strength 
by about 10~k$_\text{B}$T.  However, already in this relatively simple case many interactions
are present. First, we have the osmotic and vdW interactions $w_\text{excl+vdW}(z)$. 
Secondly,  we have the electrostatic term $w_{\rm elec}(z)$ with the repulsive monopole term but the attractive 
(monopole) Born term.  We can neglect the influence of the very small dipole of this CPPM ($P^0_0$) and counterion-release plays
no role as no patches are present.  The phenomenological prediction $w_\text{osm}(z)+w_{\rm elec}(z)\simeq w_{\rm tot}(z)$  is also 
displayed in the figures. Here, the contribution for $w_\text{osm}(z)$ was taken 
directly from the previous reference simulations of the neutral sphere in the charged brush, cf. Fig.~3(b). 
For the electrostatic contributions, eq.~(5)--(7), the values for $\Phi(z)$, $E(z)$, and charge profiles $c_i(z)$  
were taken from the simulations of an isolated charged brush, cf. Fig. 2.   The comparison is satisfying given 
the approximations  already made. 

\subsection{Interaction between a PE brush and one-patched CPPMs}

\begin{figure}[t!]
 \centering
 \includegraphics[width=0.4\textwidth]{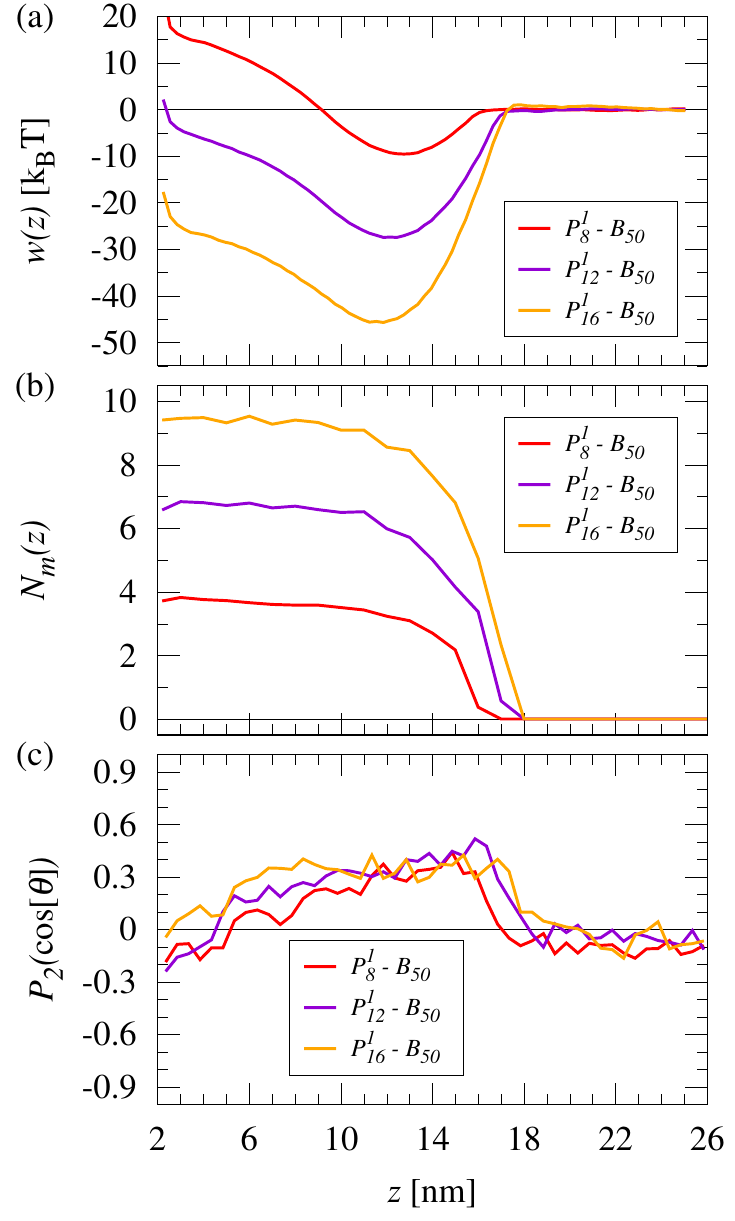}
 \caption{(a) Simulated PMF profiles $w(z)$ for the ($P^1_s$) system with $s=8,12,16$ at $c_s=32$~mM.
  (b) Number of accumulated monomers $N_m(z)$ on the patch as a function of $z$. 
   (c) CPPM orientation with respect to the grafting surface.}
 \label{fig:cppm_brush_one_patched_20mM}
\end{figure}

We now turn to the PMFs between the PE brush and a like-charged CPPM with only one patch, i.e., ($P^1_s$). 
We particularly investigate how the patch charge and ionic strength affects the PMF and its individual contributions.  
In Fig.~\ref{fig:cppm_brush_one_patched_20mM}~(a) we
present PMFs for patchiness $s=8, 12$, and 16 at ionic strengths of 32~mM. A clearly stable adsorption minimum is found at the brush surface layer at 
$z \approx 12$~nm.  By increasing the patch size from $s=8$ to $s=16$ the attraction shifts from $-10$~k$_\text{B}$T to $-45$~k$_\text{B}$T 
for $c_s=32$~mM.  

\begin{figure}[t]
 \centering
 \includegraphics[width=0.45\textwidth]{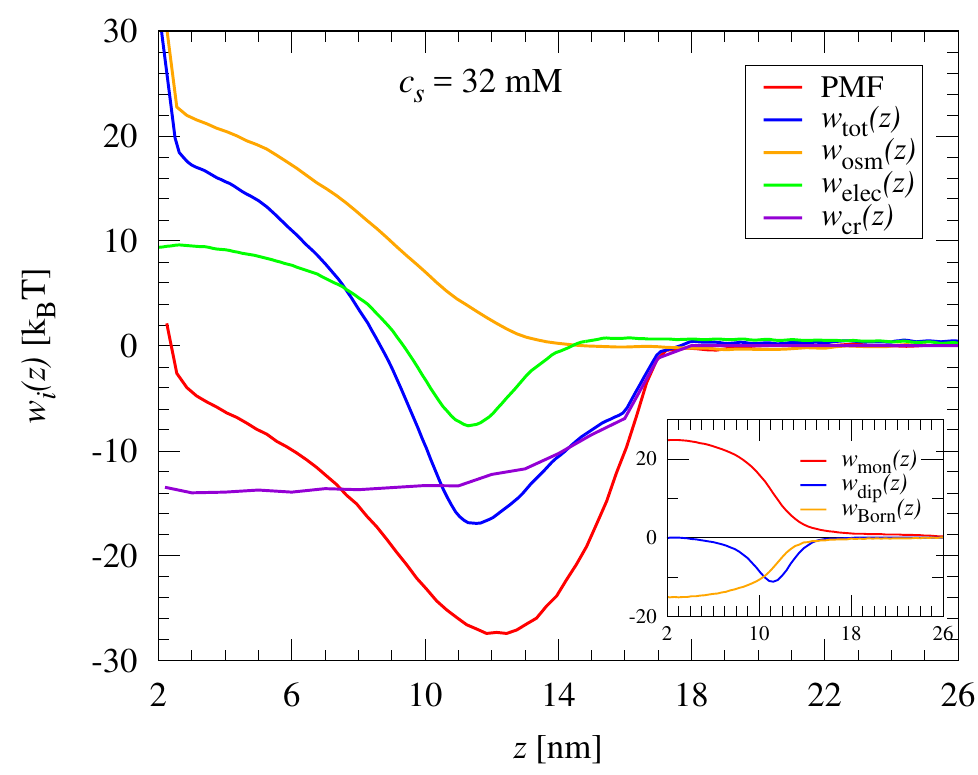}
 \caption{Individual free energy contributions to $w_\text{tot}(z)$ of the ($P^1_{12}$) complex at
  $c_s=32$~mM from eq.~(4). The insets shows the monopole (solid line), diopole (dashed line), 
   and Born (dotted) contributions to the electrostatic interaction $w_{\rm elec}(z)$ in eq.~(\ref{eq:W_elec}).}
 \label{fig:cppm_brush_individual_energies_P-1-12}
\end{figure}

The released ions are determined from the localization of $N_m(z)$ PE monomers on the positive patch and are presented in Fig.~\ref{fig:cppm_brush_one_patched_20mM}~(b).  Once the CPPM comes close to the brush surface, cf. Fig.~1(b), PE monomers interact with the positive patch 
and lead to a quick rise of the number of attached monomers. For even closer approach,  $N_m(z)$ saturates to a constant value in
the brush layer. The value depends on the patchiness and is about 3.8 and 9.3 (for $s=8$ and 16 at $z=2$~nm, respectively) PE monomers attached to 
the patch, which then leads to $N_+(z) = \left\{1-\Gamma^{-1}\right\}N_{\rm m}(z) \simeq 0.44N_{\rm m}(z)$ locally 
released counterions on average, i.e., about 1.7 and 4.1 (for $s=8$ and 16, respectively) deep in the brush.    

In Fig.~\ref{fig:cppm_brush_one_patched_20mM}~(c) the mean
angular orientation of the patch towards the grafting surfaces is shown. At large distances there is no favorite alignment of the CPPM while at intermediate distances
around $z=16$~nm, close to the brush surface, the patch vector is aligned almost parallel to the grafting surface. The alignment is expected
due to the coupling of the dipole to the $E$-field, but interestingly the maximum alignment
happens for  a distance larger than the minimum of the electrostatic field, cf. Fig.~\ref{fig:isolated_pe_brush}. The reason is 
probably that single PE chains can reach out of the brush to touch the patch and thereby strongly orient the CPPM,  see also
the snapshots in Fig.~\ref{fig:protein_uptake}(b) for an exemplary illustration. When the CPPM penetrates into the brush layer,
the orientation correlation weakens likely due to the PEs more homogeneously surrounding the CPPM and finally due to 
the decrease of the electrostatic field within the PE brush.

\begin{table*}[t!]
\centering
 \caption{A summary of the salt dependent energy contributions to the binding affinity of the ($P^1_{12}$) complex. Here, $w_{\rm min}=w(z_{\rm min})$ is the value of the simulation PMF evaluated at the global minimum $z_{\rm min}$. $w_{\rm tot}$ is evaluated from  the phenomenological description eq.~(4), where the individual terms are calculated from the data (potential and ion distributions) in the reference simulations in Figs.~2 and 3. }
 \begin{tabular}{|c|c|c||c|c|ccc|cc|c|}\hline
  $c_s$ & $z_{\rm min}$ & $w_\text{min}$ & $w_\text{excl+vdW}$ & $w_\text{elec}$ & $w_{\rm mon}$       & $w_{\rm dip}$       & $w_\text{Born}$ & \multirow{2}{*}{$N_+$} & $w_\text{cr}$ & $w_\text{tot}$\\
  $[$M] & [nm]      & k$_\text{B}$T  & $k_\text{B}$T  & k$_\text{B}$T   & k$_\text{B}$T & k$_\text{B}$T & k$_\text{B}$T   &                      & k$_\text{B}$T & k$_\text{B}$T\\\hline
  15    & 12.5     & \bf{-27.2}    & 0.7           & -5.0           & 10.0          & -10.0       & -5.0          & 2.6                 & -14.2        & \bf{-18.5}\\
  32    & 11.9     & \bf{-27.4}    & 1.3           & -6.8           & 8.7          & -10.1        & -5.4          & 2.7                 & -12.7        & \bf{-18.2}\\
  78    & 11.9     & \bf{-22.3}    & 0.6           & -4.7           & 4.0          & -6.0        & -2.8          &  2.6                 & -9.9        & \bf{-14.0}\\
  137   & 10.7     & \bf{-22.1}   & 2.0           & -5.9           & 4.3          & -6.5         & -3.7           & 2.8                 & -9.1         & \bf{-13.0}\\
  259   & 10.4     & \bf{-15.9}   & 1.3           & -3.3           & 2.2          & -3.7         & -1.9          & 2.7                 & -7.0         & \bf{-9.1}\\\hline
 \end{tabular}
 \label{tab:energy_contributions}
\end{table*}

\begin{figure}[htb]
 \centering
 \includegraphics[width=0.4\textwidth]{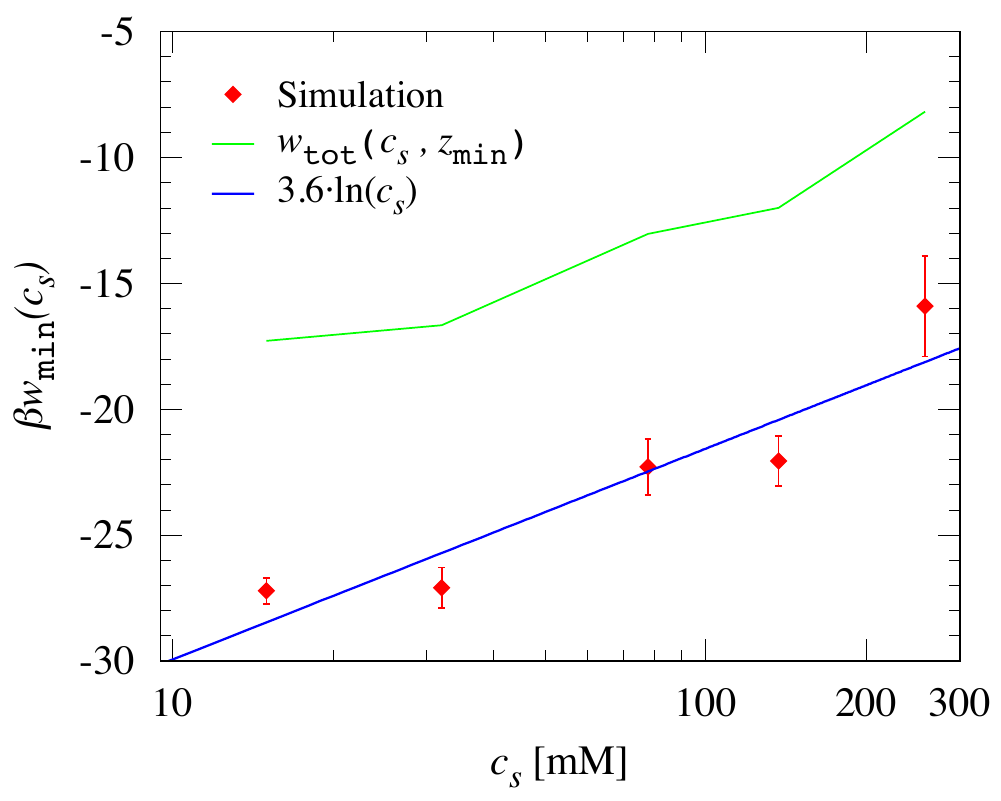}
 \caption{Binding affinity $\beta w_{\rm min}$ (minimum value of the PMFs) as a function of the ionic
 strength $c_s$ for the ($P^1_{12}$) system. The blue line is a fit according to the
 function $\beta w_{\rm min}(c_s)=\tilde{a}+\tilde{N}\ln(c_s)$ with $\tilde{a}=-36.8$ and $\tilde{N}=3.6$.
 The green line is a description by Eq.~(\ref{eq:W_tot}), see text for explanation. The individual contibution to 
 $w_{\rm min}$ are shown in Table~\ref{tab:energy_contributions}.
 }
 \label{cppm_brush_P-1-12_B-50_binding_affinity}
\end{figure}

In Fig.~\ref{fig:cppm_brush_individual_energies_P-1-12} we quantitatively compare and discuss the PMFs within 
our phenomenological framework around eq.~(\ref{eq:W_tot}) for a salt concentration $c_s=32$~mM. 
Here, we compare the simulation PMF (red curve) to the total phenomenological prediction $w_{\rm tot}(z)$ that originates 
from the sum of the vdW and excluded-volume interactions $w_\text{excl+vdW}(z)$, 
the electrostatic part $w_{\rm elec}$, and the counterion-release contribution $w_{\rm cr}(z)$. 
As discussed before, the excluded volume part is highly repulsive and completely dominated by the osmotic contributions from the counterions $w_{\rm osm}$. 
This contribution for $w_\text{osm}(z)$ (yellow curve in Fig.~\ref{fig:cppm_brush_individual_energies_P-1-12}) was taken directly 
from the reference simulations of the neutral sphere in the charged brush, cf. Fig.~3(b). 
The electrostatic part is plotted as the green curve and further subdivided into monopolar repulsion and dipole and Born attraction in the inset.
Here, the monopole repulsion is very large (up to 23~$\kB T$) and partially canceled by the large Born attraction (up to -18~$\kB T$).  
The dipole part is extremal at the brush surface at $z\simeq 12$~nm and remains an attractive contribution inside
the brush. For smaller distances $z\lesssim 10$~nm the net monopole repulsion, however, wins over it. 
To obtain the prediction of the total PMF, the counterion-release contribution has to be finally added. 
The latter is highly attractive, providing up to 10--14 $\kB T$ favorable entropy for distances smaller than $z\simeq 14$~nm. 
In total the prediction adds up to show a stable adsorption at the brush surface at $z\simeq 10-12$~nm and an increasing repulsion closer to the grafting surface. 
This description (blue curve) is in qualitative accordance with the PMF (red curve) directly extracted from the simulation. Apparently, however, the phenomenological 
description is roughly overall $\simeq 10$~$\kB T$ more repulsive. We believe this discrepancy might stem from the addition of
errors we have made in the approximations of the single contributions. In particular, our estimation of the Born free energy is probably 
the weakest of all as we performed only a simple dipolar expansion for the CPPM. The CPPM has large surface charge densities 
in direct contact with the PE chains and mobile ions and thus we believe that the Born contribution is in fact more attractive than described by eq.~(7) 
and is mostly responsible for the relatively large discrepancy of the phenomenological model. 

We finally discuss the salt concentration dependence of the binding affinity $w_{\rm min}$ represented
by the (global) minimum values $w(z_{\rm min})$ of the PMFs from the ($P^1_{12}$) systems. The data from the respective
simulations are depicted in Fig.~\ref{cppm_brush_P-1-12_B-50_binding_affinity}.
We also include a Record--Lohman type of logarithmic fit of the form $\beta w_{\rm min}(c_s) = \tilde{a}+\tilde{N}\ln[c_s]$,
which yields the number of released counterions in an ion-release dominated scenario~\cite{Record1976:1}.
From the best fit (blue line) we get for $\tilde{a}=-36.8$ and for $\tilde{N}=3.6$. The latter indicates ion release in
the order of 3 to 4 counterions as the CPPM adheres to the brush surface. Although the adsorption is not entirely driven by counterion release, this number is not too far from what we actually observe in our simulations, where we calculate roughly 2.7 ions on average, see the $N_+$ values in Table~\ref{tab:energy_contributions}. 
We also show the results of the phenomenological model from eq.~(\ref{eq:W_tot})  in Fig.~\ref{cppm_brush_P-1-12_B-50_binding_affinity}. 
Although this approach reveals a systematic deviation between the prediction (green line) and simulated
data (red points), it qualitatively reproduces the correct trend. Such a good reproduction is non-trivial as the final binding affinity results from 
cancellation of several repulsive and attractive contributions (with different individual salt concentration dependencies) 
as summarized in detail in Table~\ref{tab:energy_contributions}.


\subsection{Interaction between a PE brush and two-patched CPPMs}

\begin{figure}[t]
 \centering
 \includegraphics[width=0.4\textwidth]{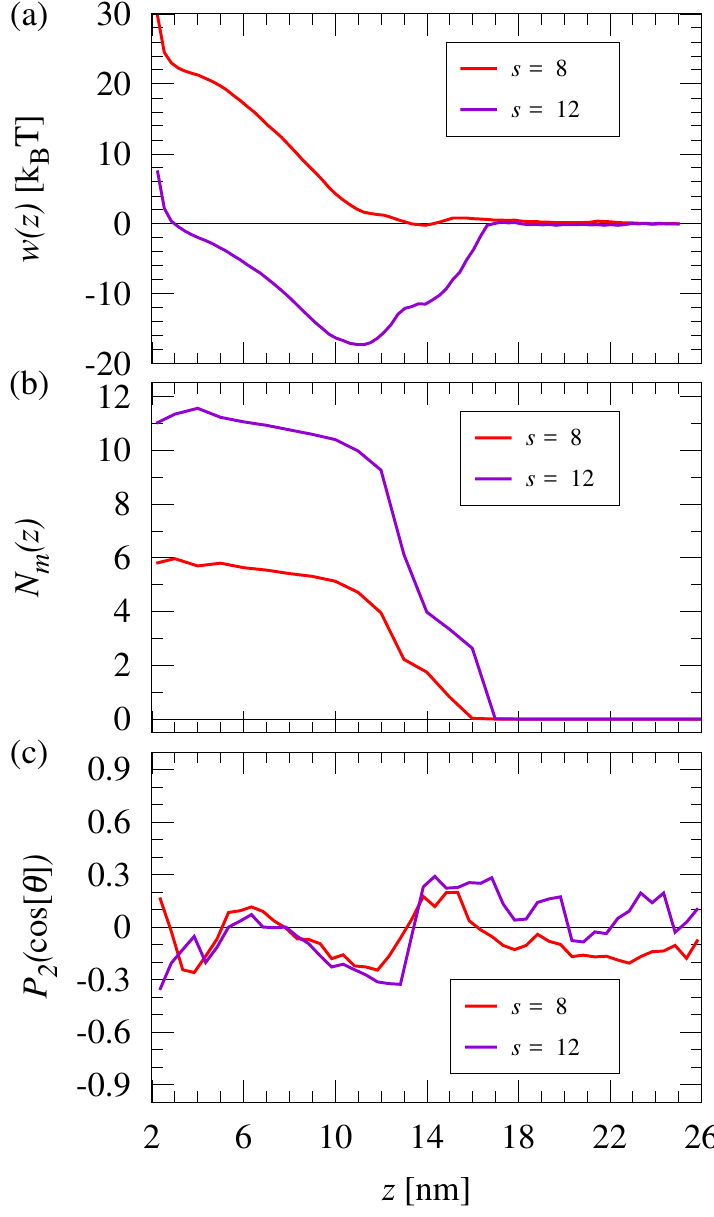}
 \caption{(a) Simulated PMF profiles of ($P^2_{s}$) with antipodally aligned patches $(m=2)$
  for patch charges $s=8$ and $s=12$ in the presence of $c_s=32$~mM salt concentration.  
  (b) Number of attached PE monomers $N_m(z)$ on the patch surface as a function of $z$. 
  (c) Patch orientation with respect to the grafting surface.}
 \label{fig:cppm_brush_two_patched_P-2-8}
\end{figure}

The manner in which a CPPM with two, antipodally aligned patches $(m=2)$ with $s=8$ and $s=12$ charges interacts with the PE brush
is presented in Fig.~\ref{fig:cppm_brush_two_patched_P-2-8}~(a) for the 32mM salt concentration. 
For $s=8$ the PMFs are essentially repulsive apart from a very shallow minimum at the brush surface at around $z\simeq 14$~nm. 
This result is actually remarkable as now two patches, 
that is, twice the number of positive charges, are acting as attractors. Recall that with only one patch, see the PMF of P$_8^1$ in 
Fig.~\ref{fig:cppm_brush_one_patched_20mM}~(a), we found a significant attraction of $\simeq 9$~$k_{\rm B}T$.  
In light of our phenomenological theory, this implies that the increased (attractive) counterion-release effects are fully compensated by 
the monopole and osmotic repulsions,  while the attractive dipolar (cf. Table~1) (and higher order multipolar) 
 interactions as well as the Born energy have become 
apparently too small to significantly stabilize the minimum.  The situation changes in the case of  larger patch charges, $s=12$, as also shown in 
 Fig.~\ref{fig:cppm_brush_two_patched_P-2-8}~(a). Here, evidently the attractive terms, presumably most 
 significantly the counterion-release contribution, clearly outweigh the repulsive interactions.  
The amount of released counterions is about a factor of two stronger in the $s=12$ system than for $s=8$, as indicated by the 
 number of adsorbed monomers shown in Figs.~\ref{fig:cppm_brush_two_patched_P-2-8}~(b).

Note also that there is an interesting small plateau-like region in the PMF of the P$_{12}^2$ CPPM 
at around $z\simeq 14$~nm separated by a small hump ($z\simeq 13$~nm) to the stable minimum at $z\simeq 11$~nm.   
We found a similar, but more pronounced behavior previously in the PMFs between two-patched CPPMs and a single 
PE~\cite{Yigit2015:1}. The reason in the latter case is  that after the first, favorable contact of the PE monomers with one of the two positive patches, 
a tighter complexation is only possible after the PE wraps around the CPPM to reach and adsorb to the second patch, eventually stabilizing the final complex. 
However, this wrapping is penalized by electrostatic repulsion (as the PE moves along the like-charged part between the patches of the globule)  and an energy barrier  can appear. It seems a similar signature can be observed here by two-patched CPPMs interacting with PE brushes. 
Noteworthy, the experimental adsorption of $\beta$-lactoglobulin to like-charged  
spherical PE brushes proceeded in a yet unexplained two-step process~\cite{Henzler2010:1}.

Finally, the local CPPM orientation, plotted in Fig.~\ref{fig:cppm_brush_two_patched_P-2-8}~(c), 
exhibits a rather complex behavior with several extrema, especially for the higher charged, $s=12$, system. 
One maximally oriented configuration of aligned patches parallel to the $z$-direction 
appears at the brush surface at $z\approx14$~nm, when the first patch interacts with the PE ends.  This is understandable again by a structural picture of PEs reaching out towards the first  approaching patch and forcing rotation of this CPPM patch to face the brush. The explanation of the appearance of the other, smaller maximum at $z\simeq 6$~nm,  or the two minima at $z\simeq 3$--$4$ and $12$~nm,  where the axis connecting the patches is perpendicular to $z$, however, is not so simple.  In general, quadrupolar interactions with electric field gradients and orientation-dependent patch adsorption to the inhomogeneously (in $z$-direction) distributed PE 
monomers rule the game, probably a mix thereof, and we leave the detailed exploration of these interesting structural phenomena of multipolar  adsorption for future studies. 

\section{Summary and concluding remarks}

In summary, we have explored the interaction of simplified patchy protein models and a thin film of a dense PE brush using  implicit-water, explicit-salt Langevin dynamics computer simulations. We focused on the regime of highly charged PE 
chains and proteins with considerable patchiness to investigate in detail the competition of osmotic and electrostatic 
interaction mechanisms. We neglected the possibilty of charge regulation and charge inversion of the protein. 
The matter of what is the driving force of like-charge attraction of proteins to PE brushes has been  controversially discussed in 
 literature~\cite{Wittemann2003:1, Biesheuvel2005:1, Wittemann2006:1, Leermakers2007:1, deVos2009:1}.

We have found that the patchy particle $P^1_8$, which possesses the lowest patch charge density
and similar electrostatic features to lactoglobulin~\cite{Ferry1941:1}, already leads to a significant adsorption 
at low and intermediate (physiological) ionic strengths. In all cases we found that the 
adsorption takes place on the surface of the PE brush, which is reflected by the location of the global PMF minimum.
Adsorption has been found to be stronger for lower salt concentrations and higher patch charge.
We note that stronger surface adsorption on the brush surface with higher particle dipole has been 
also reported by Hu \textit{et al.} who simulated a coarse-grained model of fullerene-like patchy particles~\cite{Hu2009:1}. 
 
Our analysis has demonstrated the existence of multiple competing interaction mechanisms, most notably the repulsive osmotic and monopolar electrostatic 
contributions and the attractive multipolar, self-energy (Born), and counterion-release mechanisms.  The purely electrostatic contributions due to the 
inhomogeneous charge distributions, leaving counterion-release aside, have been discussed on a global level previously in the framework of a  
mean-field Poisson--Boltzmann theory~\cite{Leermakers2007:1}. We have described the leading order parts analytically and 
found that the balance between all of those is complex and specific to the system. In our case the final adsorption was driven equally 
by electrostatic and counterion-release mechanisms, overcoming the large monopole and osmotic repulsive interactions. 
The strong affinity of the CPPM specifically to the brush surface can be traced back to the large dipole interaction, which is maximal at the brush surface, whereas the monopole and osmotic repulsions steeply rise further inside the brush and dominate over 
the  attractive contributions. 

Cleary, the attractive interactions are determined by the magnitude of the charge of the patches and thus
a minimal patch size needs to be present to compete against the repulsion~\cite{Leermakers2007:1} which, in turn, is defined by the brush charge and the protein net charge. 
Here, the counterion-release effect is a strong contributor to like-charged attraction as every condensed ion released from a PE chain provides several $\kB T$ of binding entropy.  Note that we only found these effects taking place on the PE chains, no distinctive condensation behavior could
be found on the protein patches~\cite{Yigit2015:2}.  In order to have the counterion-release effects active, the patch should probably carry
at least 2--3 localized charges, such that a PE is likely to locally bind. We have also found a non-trivial effect of the number (or geometrical distribution) 
of  patches, namely doubling the patches did not lead to a higher attraction; in contrast, the double-patched $P^2_8$ system showed 
no stable adsorption, whereas the single-patched $P^1_8$ did, probably due to the missing dipole attraction and possibly lowered self-energy contribution, 
such that the monopolar repulsion could not be overcome. So, while the distribution of patches probably has less influence on counterion
release, it has a major influence on multipolar and Born electrostatic contributions.  In that respect, also the salt concentration 
dependence of the binding affinity is non-trivial in general, as argued in our phenomenological framework,  
because all contributions have quite different functional salt dependencies. 

To conclude, the question of what is the driving force of like-charge attraction of proteins to PE brushes~\cite{Wittemann2003:1, Biesheuvel2005:1, Wittemann2006:1, Leermakers2007:1, deVos2009:1} has no unique answer; there are at least three mechanisms, and which one dominates 
is system-dependent and governed by pH, PE charges, salt concentration, and charge heterogeneity of the
protein. If, for instance, charge heterogeneity is small and experiments are operated close to the isoelectric point, charge regulation may play an important role~\cite{Biesheuvel2005:1}. Far away from the isoelectric point,  multipolar electrostatics governs the attraction. If then also the 
PE charge density is high enough, roughly beyond the Onsager-Manning-Oosawa  threshold~\cite{Fuoss1951:1, Oosawa1971:1,Manning1969:1},
the counterion-release mechanism sets in and, due to the large entropy gain per ion, can become the decisive driving force. 


\section*{Acknowledgments}
The authors thank Jan Heyda and Stefano Angioletti-Uberti for inspiring discussions. The authors are thankful for support by the Helmholtz Virtual Institute for Multifunctional Biomaterials for Medicine, and the Helmholtz Association through the Helmholtz-Portfolio Topic ``Technology and Medicine.''
 M.K. and J. D. acknowledge funding by the ERC (European Research Council) Consolidator Grant under project number 646659-NANOREACTOR.

\appendix
\section{{Appendix: Born free energy of a dipolar sphere}}
\label{app:1}

\textcolor{black}{Multipole charging free energies in dielectric and salty environments have been discussed in detail before~\cite{Kirkwood, Phillies}. Here we explicitly present the Born (self) energy for the dipole in a salty environment for the convenience of the reader. Extensions to quadrupolar and higher order terms can be performed analogously.}

We consider a point dipole $\mu$ enclosed in a spherical shell with a radius $R$. Outside the sphere we assume DH electrostatic screening $\kappa$, whereas the interior is free of ions. The corresponding electrostatic potential outside the sphere is (constructed by Eq.~(18) in Hoffmann {\it et al.}~\cite{Hoffmann}),
\begin{eqnarray}
\phi(r,\theta)&=&\frac{3\mu}{4\pi\varepsilon\varepsilon_0 \kappa R^3 } \left(\frac{R}{r}\right)^{1/2}\frac{K_{3/2}(\kappa r)}{K_{5/2}(\kappa R)}\,\cos\theta\\
&=&\frac{3\mu (1+\kappa r)}{4\pi\varepsilon\varepsilon_0 r^2 (3+3\kappa R+\kappa^2R^2)}\,\rme^{-\kappa(r-R)}\cos\theta.\nonumber
\label{eq:phi}
\end{eqnarray}
We now calculate the charging (Born) free energy outside the sphere. 
In general, the electrostatic free energy can be evaluated as
\begin{equation}
w_{\rm Born}=\frac{1}{2}\int \rho(\Av r)\phi(\Av r)\rmd \Av r.
\end{equation}
Since we want to evaluate only the free energy outside the sphere, we cannot use the actual charge distribution $\rho(\Av r)$. 
We can transform the above integral using linearized PB equation $(\nabla^2-\kappa^2)\phi=-(1/\varepsilon\varepsilon_0) \rho(\Av r)$, and thus eliminate the density,
\begin{equation}
w_{\rm Born}=-\frac{1}{2}\varepsilon\varepsilon_0\int \phi(\Av r)\nabla^2 \phi(\Av r)\rmd \Av r
+\frac{1}{2}\varepsilon\varepsilon_0\kappa^2\int \phi^2(\Av r)\rmd \Av r
\end{equation}
Now we use a relation $\nabla^2\phi^2=2(\nabla\phi)^2+2\phi\nabla^2\phi$. It can be shown that $\int\nabla^2\phi^2\rmd\Av r=0$, and therefore we can rewrite the free energy as
\begin{equation}
w_{\rm Born}=\frac{1}{2}\varepsilon\varepsilon_0\int (\nabla\phi)^2\rmd \Av r +\frac{1}{2}\varepsilon\varepsilon_0\kappa^2\int \phi^2(\Av r)\rmd \Av r.
\label{eq:F}
\end{equation}
The above free energy expression has a simple physical interpretation; the first term is the integral over energy density of the electric field  $(1/2)\varepsilon\varepsilon_0 E^2$, and the second term stems from the entropy of ions, $\kB  (n_+\log n_++n_-\log n_-)$. The term $(\nabla \phi)^2$ expresses in spherical coordinates  as
\begin{equation}
(\nabla\phi)^2=\left(\frac{\partial\phi}{\partial r}\right)^2+\frac{1}{r^2}
\left(\frac{\partial\phi}{\partial \theta}\right)^2.
\end{equation}
We insert the potential $\phi(r,\theta)$ given by Eq.~(\ref{eq:phi}) into Eq.~(\ref{eq:F}) and integrate over $r\in (R,\infty)$ and $\theta\in(0,\pi)$, which yields
\begin{equation}
w_{\rm Born}(\kappa)=\frac{3\mu^2}{8\pi\varepsilon\varepsilon_0 R^3}\,\frac{(1+\kappa R)[2+2\kappa R+(\kappa R)^2]}{[3+3\kappa R+(\kappa R)^2]^2}.
\label{eq:F_born_dipole}
\end{equation}
\textcolor{black}{If the sphere has both a monopole $Q$ and dipole $\mu$ the total Born term is given by eq.~(6). One can show  that the cross monopole-dipole term vanishes in the above linear treatment.}

\clearpage 
\providecommand{\latin}[1]{#1}
\providecommand*\mcitethebibliography{\thebibliography}
\csname @ifundefined\endcsname{endmcitethebibliography}
  {\let\endmcitethebibliography\endthebibliography}{}

\end{document}